\documentclass[a4paper, twoside, 11pt]{article}

\usepackage[a4paper,rmargin=3.5cm,lmargin=3.5cm, twoside]{geometry}
\usepackage{amssymb}
\usepackage{amsmath}
\usepackage[numbers,sort&compress]{natbib}
\usepackage{epsfig}
\usepackage{bbm}

\newcommand{\mtin}[1]{\mbox{\tiny {#1}}}
\newcommand{\ca}[1]{{\cal #1}}
\newcommand{\ol}[1]{\overline{#1}}
\newcommand{\sfrac}[2]{{\textstyle\frac{#1}{#2}}}

\begin{document}
\bibliographystyle{amsplainchanged}
\begin{center}
{\bf\LARGE\large Efficient Parametrization of the Vertex Function, $\Omega$--Scheme, and the $(t,t')$--Hubbard
Model at Van Hove Filling}
\end{center}
\renewcommand{\thefootnote}{\fnsymbol{footnote}}

\begin{center}
Christoph Husemann\footnote{c.husemann@thphys.uni-heidelberg.de} and Manfred Salmhofer\footnote{m.salmhofer@thphys.uni-heidelberg.de}\\ 
{\em Institut f\"ur Theoretische Physik, Universit\"at Heidelberg\\ D-69124 Heidelberg, Germany} \\
(January 12, 2009)
\end{center}

\renewcommand{\thefootnote}{\arabic{footnote}}
\begin{abstract}
We propose a new parametrization of the four--point vertex function
in the one--loop one--particle irreducible renormalization group (RG)
scheme for fermions. 
It is based on a decomposition of the effective two--fermion
interaction into fermion bilinears that interact via exchange
bosons. The numerical computation of the RG flow of the boson
propagators reproduces the leading weak coupling instabilities of
the two--dimensional Hubbard model at Van Hove filling, as they were
obtained by a temperature RG flow in \cite{TemperatureFlow}. Instead
of regularizing with temperature, we here use a soft frequency
$\Omega$--regularization that likewise does not artificially suppress
ferromagnetism. Besides being more efficient than
previous $N$--patch schemes, this parametrization also reduces the
ambiguities in introducing boson fields.
\end{abstract}

\setcounter{footnote}{0}
\section{Introduction}
In the past decade fermionic Wilsonian renormalization group (RG)
methods have been very successful in classifying weak coupling
instabilities of the two--dimensional $(t,t')$--Hubbard model \cite{ZanchiSchulz1,ZanchiSchulz2000,HalbothMetzner,HalbothMetznerPomeranchuk,NpatchRG,Umklapp,TemperatureFlow}. 
The starting point in these methods is an exact functional equation.
Besides being very useful for proving mathematical statements, this
equation can be approximated to allow a direct numerical
calculation. In the case of weak coupling an expansion in the
fermionic fields can be truncated by setting the six--point function
to zero. For an appropriate Fermi surface geometry this one--loop
approximation is also justified for intermediate couplings and scales \cite{salmfrgtt}.

Successes of the fermionic one--loop RG method include the
explanation of the interplay of antiferromagnetic and
superconducting instabilities in the Hubbard model for small next to
nearest neighbor hopping $-t'$. Results were first
obtained using Polchinskis version of the RG
\cite{ZanchiSchulz1,ZanchiSchulz2000}, 
the Wick ordered scheme
\cite{HalbothMetzner,HalbothMetznerPomeranchuk}, and the
one--particle irreducible (1PI) scheme \cite{NpatchRG,
  Umklapp}.
Near half filling, where the Fermi surface is nested,
antiferromagnetism was found to be the leading instability. For
smaller fillings on the hole doped side, where the Fermi surface is more strongly curved and regular, $d_{x^2-y^2}$--wave superconductivity is dominating
(although not present initially in the repulsive Hubbard
model, such a term is generated as an effective interaction 
in the RG flow; specifically, it is induced by antiferromagnetic
correlations). 
In between the two regions the antiferromagnetic and
superconducting tendencies mutually reinforce each other
\cite{Umklapp,HonerkampSelfEnergy}. This saddle point regime,
where the main contribution comes from the saddle or Van Hove
points, is interpreted as the weak coupling analogue of the Mott
state, but not yet fully understood.

In all works mentioned in the last paragraph a Fermi surface
momentum or frequency cut--off was used. Although being
conceptionally clear, this regularization suppresses
small--momentum particle--hole fluctuations, 
as pointed out in \cite{TemperatureFlow}. 
Overcoming this drawback by using temperature
as a flow parameter, 
the leading instability for larger hopping $-t'\in[0.34t,0.5t]$
was found to be ferromagnetism at Van Hove filling and triplet
superconductivity away from Van Hove filling \cite{TemperatureFlow,
TemperatureFlow2}. This result was qualitatively confirmed by the
same method in \cite{katanin2patch}, by an interaction RG flow
\cite{gflow}, and a two--particle self--consistent Monte Carlo
approach \cite{Tremblay}. The mutual suppression of ferromagnetic
and $d$--wave superconducting tendencies decreases the critical
scale at Van Hove filling for intermediate $-t'$ by several orders
of magnitude. Within numerical accuracy, 
the instability analysis of \cite{TemperatureFlow,
TemperatureFlow2} even suggests the existence of a
quantum critical point between the ferromagnet and the
superconductor. 

The one--loop 1PI RG flow used here is described by the
evolution of an effective four--point vertex function and a
selfenergy. Even if the selfenergy is not taken into account, the
flow equation for the four--point function is a non--linear
integro--differential equation (see Section
\ref{sec:RenormalizationGroup}). This equation has to be solved for a
function with three independent momenta and three independent
frequencies. For a numerical computation, as applied by the above
mentioned studies, further approximations are needed. By Taylor
expansion and power counting it can be argued that the main
contribution to low energy excitations comes from frequency zero.
All above mentioned one--loop studies neglect the frequency
dependence of the vertex function and evaluate the right hand side
of the flow equation at frequency zero. In a next step the momentum
dependence is discretized. Again by a low energy argument, the
momenta of the vertex function are projected onto the Fermi surface
in most one--loop studies. In a so--called $N$--patch scheme 
\cite{FMRTInfiniteVolume,ZanchiSchulz1} the Fermi surface is
divided into $N$ patches,\footnote{This only applies for $N>2$. The extreme case are the two--patch models, where, contrary to $N$--patch schemes, full momentum
space is not taken into account. At Van Hove filling they expect the main
contribution to come from two small patches around the
Van Hove points \cite{Salm2patch,Umklapp,katanin2patch}.} and the angular dependence of the vertex 
on each momentum is approximated by a constant in each patch. Evaluating
integrals as sums over patches, this leads to a coupled system of
$\sim N^3$ ordinary, non--linear differential equations, which can
be solved numerically. In \cite{gflow} the momentum dependence is
not projected onto the Fermi surface, so that full momentum space
$(-\pi,\pi]^2$ has to be covered by patches.

Despite the successes of the one--loop $N$--patch schemes, the
method shows room for improvement. 
-- First of all, the selfenergy is
neglected in most one--loop RG studies. Quasi particle scattering
rates of the Landau Fermi liquid are calculated in
\cite{HonerkampSelfEnergy,
MetznerSelfEnergy,ZanchiSelfEnergy,NpatchRG,
KataninKampfSelfEnergy}, which gives a first hint of selfenergy
effects. Deformations of the Fermi surface due to the real part of
the selfenergy are taken into account in \cite{Umklapp}. A
new dynamical adjustment of the propagator \cite{SalmAdjustment} can
help to include the selfenergy entirely. 
-- A second problem is the observed flow to strong coupling, 
which limits the validity of the one--loop RG flow. The flow 
has to be stopped before all scales are 
integrated out. That is, the full model is not recovered in the flow
since the infrared regularization is not removed completely. This
drawback may be overcome with a modified one--loop scheme by
allowing the flow to continue in a symmetry broken phase
\cite{symmetrybrokenBCS,KataninWard}. Using this modified scheme the
exact solution of a mean--field model for superconductivity is
obtained in the thermodynamic limit
\cite{symmetrybrokenBCS,EliashbergBCS}. Furthermore, the scheme has
been recently applied to the attractive Hubbard model
\cite{attractiveHubbard}. Due to the emergence of an effective
superconducting gap, which regularizes infrared divergencies, all
scales can be integrated out. However, since the vertex function is
not charge invariant anymore, the flow equations become very
complicated. 
Besides the purely fermionic description of the RG
flow, a bosonic continuation with mean--field theory
\cite{RGMeanField} or partially bosonized RG flows
\cite{MetznerPartial,WetterichKrahl,WetterichAFOrder,Kopietz} are promising
methods to treat the symmetry--broken phases.

We also address a third problem of the one--loop $N$--patch schemes,
which is the parametrization of the effective vertex function.
Although the initial four--point vertex function of the Hubbard
model is a constant in momentum space, a 
nontrivial and physically important momentum
structure evolves in the one--loop RG flow. 
The approximations entailed by the projections in $N$--patch schemes are not yet fully understood.
A severe restriction on the possible choice of $N$ is the high computational cost
of solving $\sim N^3$ ordinary differential equations, whose
coefficients have to be calculated as two--dimensional integrals in
every integration step. For the same reason, including the frequency dependence is very costly. 
A more efficient parametrization of the four--point vertex function should
identify the relevant processes. Their separation from irrelevant
remainders can be guided by previous one--loop studies and the
observation that the identification of the leading instabilities is determined by the
asymptotic singular structure of the flow equation. 

In Section \ref{sec:decomposition} we propose a new parametrization
of the vertex function in the one--particle irreducible RG scheme.
Guided by the singular momentum structure of the right hand side of
the flow equation, we identify three channels with distinct singular
momentum structures.\footnote{In a similar approach the effective vertex function of the single impurity Anderson model was recently written as a sum of three functions, each dependent on only one singular frequency \cite{Karrasch}.} All graphs of the flow equation are uniquely
assigned to a channel and the channels are general two--fermion interactions, so this is no approximation. However, since
we define the channels based on their singular momentum structure,
there are still ambiguities involved. This condition allows
different definitions, for example, how a constant term is
distributed over the channels. Therefore the initial on--site interaction is
not assigned to a channel but kept fixed in the flow explicitly.
Nevertheless, the definition of the channels in terms of graphs
is natural from the structure of the flow equation. 

In the thus defined channels the effective two--fermion interaction is expanded
in terms corresponding to fermion bilinears that interact via exchange bosons. 
For this study, we choose scale-- and frequency--independent form factors for the fermion  bilinears. 
In Section \ref{sec:superconductivity}, we show that
only a small number of terms is needed to capture the essential
features of the one--loop RG flow in the case where one channel is dominant. In the case of competing channels, we choose the parametrization guided by the results from $N$--patch studies. Inserting this decomposition into
the RG equation and projecting the right hand side onto the
coefficients of the expansions, the flow equations for the boson
propagators are derived. In Section \ref{sec:RGequations} we state
these flow equations for general expansions with a finite number of form factors
when the tails of the expansions are neglected.

We test the proposed parametrization of the vertex function in an
application to the $(t,t')$--Hubbard model at Van Hove filling and
temperature zero. Because at this filling, the interplay between superconductivity and ferromagnetism is important, 
it is essential to choose a scale--dependent regularization
that does not artificially suppress small momentum particle--hole
fluctuations \cite{TemperatureFlow}.
Here we do not use the temperature flow, but instead 
choose a very mild infrared regulator that keeps these 
fluctuations, but provides enough regularization to have
a well--defined flow equation, see Section \ref{sec:regularization}.
Further approximations are applied for the numerical implementation in Section \ref{sec:patching}, where we neglect the frequency dependence
of the boson propagators and discretize their momentum dependence using step functions. Since this momentum dependence is mainly determined by the one--loop bubbles, there is
good guidance in that. Also, we choose a small number of form
factors. The numerical results are stated and discussed in Section 
\ref{sec:results}. We obtain the same ordering tendencies of the Hubbard model at Van Hove filling as in \cite{TemperatureFlow}.

In the remaining part of this introduction we describe the
model and our notations. Consider the
two--dimensional quadratic lattice
$\Gamma=\mathbb{Z}^2/L\mathbb{Z}^2$ with unit spacing, length
$L\in2\mathbb{N}$, and periodic boundary conditions. Then the dual
lattice or momentum space is given by the torus $\Gamma^* =
\frac{2\pi}{L} \mathbb{Z}^2 / (2\pi \mathbb{Z}^2)$. 
Let $a_{\mathbf{p},\sigma}$ and
$a^{\dagger}_{\mathbf{p},\sigma}$ be the fermion operators on the
associated Fock space with momenta $\mathbf{p}\in\Gamma^*$ and spin
$\sigma\in \{+,-\}$ in quantization direction. The $(t,t')$--Hubbard
model on $\Gamma$ is defined by the Hamiltonian
\begin{align}\label{eq:Hamiltonian}
\ca{H}[a^{\dagger},a]=\sum_{\mathbf{p}\in\Gamma^*
\atop\sigma\in\{+,-\}} e(\mathbf{p})
a_{\mathbf{p},\sigma}^{\dagger} a_{\mathbf{p},\sigma} + U
\sum_{\mathbf{p_1} \ldots \mathbf{p_3}\in\Gamma^*}
a_{\mathbf{p_1},+}^{\dagger} a_{\mathbf{p_2},-}^{\dagger}
a_{\mathbf{p_3},-} a_{\mathbf{p_1}+\mathbf{p_2}-\mathbf{p_3},+}
\end{align}
with an on--site repulsion $U>0$ between electrons. The kinetic term of
(\ref{eq:Hamiltonian}) emerges in a tight binding approximation.
That is, the transfer integral, which describes electron hopping
between different lattice sites in real space $\Gamma$, is set to $t$
for nearest neighbors and $-t'$ for next to nearest neighbors. In
momentum space this corresponds to the dispersion relation
\begin{align}\label{eq:dispersion}
e(\mathbf{p}) = -2t(\cos p_x+\cos p_y) -4t' \cos p_x \cos p_y -\mu
\, ,
\end{align}
where the chemical potential $\mu$ is already included in the
definition of $e(\mathbf{p})$ and $\ca{H}$, which is convenient for
setting up the grand canonical partition function.

In the following $t$ is set to one, that is, all quantities are
measured in energy units of $t$. The next to nearest neighbor
hopping is considered in  the parameter range $-t'\in
(0,\frac{t}{2})$. It determines the shape of the non--interacting
Fermi surface together with the chemical potential. For $\mu=4t'$
the system is at Van Hove filling where the Fermi surface intersects
the Van Hove points, where the gradient of the dispersion
(\ref{eq:dispersion}) vanishes. This causes a logarithmic divergence
in the density of states.

\section{The Fermionic 1PI One--Loop RG\label{sec:RenormalizationGroup}}
We consider the one--particle irreducible (1PI) fermionic
renormalization group (RG) flow in the one--loop approximation
\cite{salmfrgtt}. Since we restrict the RG
flow to the symmetric phase, the effective interaction, which depends
on the RG scale $\Lambda$, is assumed to be
$SU(2)$-- and $U(1)$--symmetric, as is the initial Hubbard
interaction. An effective
two--fermion interaction with those symmetries can be generally
written in terms of a four--point vertex function $V$ in the
following way
\begin{align}\label{eq:vertexfunction}
\ca{V}_{\Lambda}[\Psi]&= \frac 12 \int \mathrm{d}p_1\ldots \mathrm{d}p_4
\;\delta(p_1+p_2-p_3-p_4)\; V_{\Lambda}(p_1,p_2,p_3) \\ \nonumber
&\hspace{5cm} \times \sum_{\sigma,\sigma'\atop\in\{+,-\}}
\ol{\psi}_{\sigma}(p_1) \ol{\psi}_{\sigma'} (p_2)
\psi_{\sigma'}(p_3) \psi_{\sigma}(p_4) \, .
\end{align}
Due to momentum conservation, the four--point vertex function
$V_{\Lambda}$ is a function of three independent momenta and
frequencies. Here we denote $p=(p_0,\mathbf{p})$ with Matsubra frequency
$p_0$ and spatial momentum $\mathbf{p}$. The integral $\int \mathrm{d}p$ is
shorthand notation for $\frac{1}{\beta}\sum_{p_0}
\int_{(-\pi,\pi]^2}\frac{\mathrm{d}^2\mathbf{p}}{(2\pi)^2}$ in the
thermodynamic limit $L \to \infty$ and at inverse temperature $\beta$.
The initial on--site interaction of the
$(t,t')$--Hubbard model is characterized by the constant vertex
function $V_{\mtin{H.M.}}(p_1,p_2,p_3)=U$.

The flow equation for the
effective vertex function in the 1PI RG scheme reads
\begin{align}\label{eq:VertexFlow}
\dot{V}_{\Lambda}(p_1,\ldots p_3) &= \ca{T}_{\mtin{pp}}(p_1,\ldots p_3)+
\ca{T}_{\mtin{ph}}^{\mtin{d}}(p_1,\ldots p_3) +
\ca{T}_{\mtin{ph}}^{\mtin{cr}}(p_1,\ldots p_3) \, ,
\end{align}
where the dot denotes the derivative with respect to the scale
$\Lambda$. The particle--particle contribution and the crossed and
direct particle--hole contributions are respectively given by
\begin{align*}
\ca{T}_{\mtin{pp}}(p_1,\ldots p_3) &= -\int \mathrm{d}p
\left[{\textstyle \frac{d}{d\Lambda}} G(p) G(p_1+p_2-p)\right] V_{\Lambda}(p_1,p_2,p)\,
V_{\Lambda}(p_1+p_2-p,p,p_3)\\
\ca{T}_{\mtin{ph}}^{\mtin{cr}} (p_1,\ldots p_3) &= -\int \mathrm{d}p
\left[{\textstyle\frac{d}{d\Lambda}} G(p) G(p+p_3-p_1)\right]
V_{\Lambda}(p_1,p+p_3-p_1,p_3)\\ &\hspace{7cm}\times \, V_{\Lambda}(p,p_2,p+p_3-p_1)\\
\ca{T}_{\mtin{ph}}^{\mtin{d}} (p_1,\ldots p_3) &= \int \mathrm{d}p
\left[{\textstyle\frac{d}{d\Lambda}} G(p) G(p+p_2-p_3)\right] \Big[
2V_{\Lambda}(p_1,p+p_2-p_3,p) \,V_{\Lambda}(p,p_2,p_3) \\
&\hspace{3cm} -V_{\Lambda}(p_1,p+p_2-p_3,p_1+p_2-p_3)\,V_{\Lambda}(p,p_2,p_3)
\\& \hspace{3cm}-V_{\Lambda}(p_1,p+p_2-p_3,p)\,
V_{\Lambda}(p,p_2,p+p_2-p_3)\Big] \, .
\end{align*}
The scale derivatives in the integrands act only in the square
brackets. Since we neglect the selfenergy, the propagator $G(p)=[ip_0
+ e(\mathbf{p})]^{-1} \chi_{\Lambda}(p)$ is the free propagator
multiplied with an appropriate cutoff function. That is, although not denoted explicitely, it is understood throughout that the propagator $G(p)$ depends on scale $\Lambda$. This flow equation is
derived in \cite{salmfrgtt}.

Since the initial vertex function $U$ of the Hubbard model obeys the
following symmetries, the flow equation implies that the effective
vertex function $V_{\Lambda}$ satisfies
\begin{align}\label{eq:VSymmetry}
\begin{array}{cl}
V_{\Lambda}(p_1,p_2,p_3)=V_{\Lambda}(p_3,p_1+p_2-p_3,p_1) \quad& \mbox{(PHS)}\\
V_{\Lambda}(p_1,p_2,p_3)=V_{\Lambda}(p_2,p_1,p_1+p_2-p_3) \quad& \mbox{(RAS)}\, .
\end{array}
\end{align}
The first symmetry expresses invariance of the
partition function as a functional integral under a change of
integration variables $\psi\mapsto i\ol{\psi}$ and $\ol{\psi}\mapsto
i\psi$ and is called particle hole symmetry (PHS). The second symmetry
is directly inherited from
(\ref{eq:vertexfunction}) and accounts for remnants of the antisymmetry
(RAS) of the Grassmann variables.

\section{Decomposition of the Effective
  Interaction\label{sec:decomposition}}

The parametrization of the vertex function developed in this section
is based on the observation that mainly the singular momentum
structure of the RG equation determines the qualitative instabilities
of the flow. That is, the parametrization of $V$ should simplify the 
momentum dependence but should keep track of all possible singular
contributions. If the vertex function is regular, the only momentum
dependence that can change the singular behavior of the RG equation
is the transfer momentum that propagates through the
scale--derivative of the particle--particle or particle--hole bubble
\begin{align}
\dot{\Phi}_{\mtin{pp/ph}}(l) = \frac{\mathrm{d}}{\mathrm{d}\Lambda}
\int \mathrm{d} p \; G(p) G(l\mp p) \, .
\end{align}
In order to absorb this momentum dependence we introduce three
additional channels corresponding to three different transfer
momenta in the RG equation respectively
\begin{align}\label{eq:decomposition}
\ca{V}_{\Lambda}[\Psi] &= \ca{V}_{\mtin{H.M.}}[\Psi] +
\ca{V}_{\mtin{SC}}^{\Lambda}[\Psi] +
\ca{V}_{\mtin{M}}^{\Lambda}[\Psi] +
\ca{V}_{\mtin{K}}^{\Lambda}[\Psi] \, .
\end{align}
The initial on--site interaction of the Hubbard model
$\ca{V}_{\mtin{H.M.}}[\Psi]$ is kept in the parametrization since
there are ambiguities in assigning it to the other channels. It will
remain constant in the flow and can be seen as a driving force. This
does not mean that the on--site term contained in the full
interaction (\ref{eq:decomposition}) remains independent of scale.
The corrections to it are absorbed as contributions to the
additional channels
\begin{align}\nonumber
\ca{V}_{\mtin{SC}}^{\Lambda}[\Psi] &= -\frac 14 \int
\mathrm{d}q\mathrm{d}q'\mathrm{d}l \;
\Phi_{\mtin{SC}}^{\Lambda}(q,q',l) \sum\limits_{J=0}^{3}
\Big(\ol{\Psi}(q)\sigma^{(J)}  \ol{\Psi} (l-q)\Big)\Big(
\Psi(q')\sigma^{(J)}  \Psi(l-q')\Big) \\
\label{eq:channelinteraction} \ca{V}_{\mtin{M}}^{\Lambda}[\Psi] &=
-\frac 14 \int \mathrm{d}q\mathrm{d}q'\mathrm{d}l \;
\Phi_{\mtin{M}}^{\Lambda}(q,q',l) \sum\limits_{j=1}^{3}
\Big(\ol{\Psi}(q)\sigma^{(j)}  {\Psi} (q+l)\Big)\Big(
\ol{\Psi}(q')\sigma^{(j)}  \Psi(q'-l)\Big) \nonumber \\
\ca{V}_{\mtin{K}}^{\Lambda}[\Psi] &= -\frac 14 \int
\mathrm{d}q\mathrm{d}q'\mathrm{d}l \;
\Phi_{\mtin{K}}^{\Lambda}(q,q',l) \Big(\ol{\Psi}(q) 
{\Psi}(q+l)\Big)\Big( \ol{\Psi}(q') \Psi(q'-l)\Big)\, ,
\end{align}
where $\sigma^{(J)}$ are the Pauli matrices for $J=1,2,3$ and
$\sigma^{(0)}$ is the two--dimensional unit matrix. Each of the
channels (\ref{eq:channelinteraction}) is a general $U(1)$ and $SU(2)$
symmetric two--fermion interaction. We
call them the superconducting, magnetic, and forward scattering
channel respectively. These names are legitimate if the functions
$\Phi_{\mtin{SC}}$, $\Phi_{\mtin{M}}$, and $\Phi_{\mtin{K}}$ are
regular in their first two momentum indices and
possibly singular only in the third (bosonic) momentum index $l$.
Note that the $\Phi$'s are zero at the beginning and are generated
in the flow. In order to define their evolution we use the standard
Fierz identity $\frac 12 \sum_{J=0}^3
\sigma_{\sigma_1\sigma_2}^{(J)}\sigma_{\sigma_3\sigma_4}^{(J)}=
\delta_{\sigma_1\sigma_4}\delta_{\sigma_2\sigma_3}$ to restate
(\ref{eq:channelinteraction}) in terms of a vertex function
\begin{align}\label{eq:decompositionvertexfunction}
V_{\Lambda}(p_1,p_2,p_3) &= U
-\Phi_{\mtin{SC}}^{\Lambda}(p_1,p_3,p_1+p_2)+
\Phi_{\mtin{M}}^{\Lambda}(p_1,p_2,p_3-p_1) \\ &\quad+ \frac 12
\Phi_{\mtin{M}}^{\Lambda}(p_1,p_2,p_2-p_3) - \frac 12
\Phi_{\mtin{K}}^{\Lambda}(p_1,p_2,p_2-p_3) \nonumber
\end{align}
and insert it into the RG equation. The strong momentum dependence
of the superconducting channel should be the sum of the two incoming
particles $p_1+p_2$ which is the transfer momentum of the
particle--particle graph. Likewise the combinations $p_3-p_1$ and
$p_2-p_3$ are the transfer momenta of the crossed and direct
particle--hole graphs respectively. So we define
\begin{align}\nonumber
\dot{\Phi}_{\mtin{SC}}^{\Lambda}(p_1,p_3,p_1+p_2)
&=-\ca{T}_{\mtin{pp}}(p_1,p_2,p_3)\\ \label{eq:channelevolution}
\dot{\Phi}_{\mtin{M}}^{\Lambda}(p_1,p_2,p_3-p_1)
&=\ca{T}_{\mtin{ph}}^{\mtin{cr}}(p_1,p_2,p_3)\\\nonumber
\dot{\Phi}_{\mtin{K}}^{\Lambda}(p_1,p_2,p_2-p_3)
&=-2\ca{T}_{\mtin{ph}}^{\mtin{d}}(p_1,p_2,p_3)
+\ca{T}_{\mtin{ph}}^{\mtin{cr}}(p_1,p_2,p_1+p_2-p_3)\,.
\end{align}

This decomposition of the interaction vertex into three channels is
exact at one--loop level and the symmetries (PHS) and (RAS) are
satisfied by each channel, that is
\begin{align}
\Phi_{\mtin{SC}}^{\Lambda}(q,q',l) &=
\Phi_{\mtin{SC}}^{\Lambda}(l-q,l-q',l) = \Phi_{\mtin{SC}}^{\Lambda}(q',q,l)\\
\nonumber \Phi_{\mtin{M/K}}^{\Lambda}(q,q',l) &=
\Phi_{\mtin{M/K}}^{\Lambda}(q',q,-l) =
\Phi_{\mtin{M/K}}^{\Lambda}(q+l,q'-l,-l) \,.
\end{align}
The first identity in each line corresponds to (RAS) and can be
directly seen from (\ref{eq:channelinteraction}), whereas the second
identity corresponds to (PHS). Both symmetries follow from the
defining flow equations (\ref{eq:channelevolution}). Therefore
(\ref{eq:decompositionvertexfunction}) satisfies (RAS) and (PHS).

In summary, the definition of the channels is chosen such that each
channel carries one critical momentum dependence. If the vertex
function is still regular, all possible singular momentum dependence
of the flow equation is absorbed by the channels. Furthermore, the
decomposition
satisfies (RAS) and (PHS). The assignment of the graphs, however,
is not unique for special momenta. For example, a constant term
can be freely distributed among the channels.
Nevertheless, we have uniquely defined the channels
assuming that the different classes of graphs with their corresponding transfer
momenta form a natural decomposition. 
This is the only definition such that each channel absorbs one singular momentum and such that (if the transfer momentum becomes singular) the channels correspond to interactions of Cooper pairs, spin operators, and density operators respectively as written in (\ref{eq:channelinteraction}). 

Next we further decompose the interaction vertex by writing each
channel as an interaction of two fermion bilinears interacting via
an exchange boson. Naturally, the critical transfer momentum,
which as a sum of two fermion momenta is a boson momentum, is chosen
to be the momentum of the boson propagator. Due to (PHS),
$\Phi_{\mtin{SC}}(q,q',l)=\Phi_{\mtin{SC}}(q',q,l)$ and we can
expand
\begin{align}\label{eq:SCexpansion}
\Phi_{\mtin{SC}}^{\Lambda}(q,q',l) = \sum_{m,n\in
\ca{I}_{\mtin{SC}}} D_{mn}(l) f_m(\sfrac
 {\mathbf{l}}{2}- \mathbf{q}) f_{n}(\sfrac{\mathbf{l}}{2}-\mathbf{q}')
 +R_{\mtin{SC}}(q,q',l) \, ,
\end{align}
such that $\ca{V}_{\mtin{SC}}[\Psi]$ describes the interaction of
Cooper pairs via the boson propagator $D(l)$ up to a remainder term
$R_{\mtin{SC}}$. Actually, $\Phi_{\mtin{SC}}$ as a function of its
first two indices can be regarded as the kernel of a Fredholm
operator. But since we are not able to determine the
scale--dependent eigenfunctions, we expand in a given
(scale--independent) finite set of form factors
$(f_n)_{n\in\ca{I}_{\mtin{SC}}}$ and leave a remainder. The form
factors are orthonormalized on full momentum space $(-\pi,\pi]^2$
(divided by $4\pi^2$) and depend only on the relative momentum of
the Cooper pairs, representing different singlet and triplet lattice
symmetries. In principle an expansion in momentum and frequency
degrees of freedom is possible. For simplicity we take an approach
where the form factors (in contrast to the boson propagator $D$) do
not depend on frequency. This leaves an ambivalence in the
definition of $D$ such that the expansion (\ref{eq:SCexpansion}) is
satisfied. We set the frequency corresponding to $q$ and $q'$ to
$\frac{l_0}{2}$, which is a fermion frequency since $l_0$ is a boson
frequency. The leading instability will occur at $l_0=0$ so in that
case $q_0=q_0'=0$ which will give the main contribution. For
$\beta<\infty$ this is an overestimate since the lowest possible
(absolute) value of a fermion frequency is $\pm \frac{\pi}{\beta}$.
Alternatively for finite temperature, $q_0$ and $q_0'$ could be set
to $\frac{l_0}{2} \pm \frac{\pi}{\beta}$ and $\frac{l_0}{2} \mp
\frac{\pi}{\beta}$ respectively, with symmetrization over the sign.
That is, for $m,n\in\ca{I}_{\mtin{SC}}$, the coefficient $D_{mn}(l)$
in (\ref{eq:SCexpansion}) is given by
\begin{align}\label{eq:Ddefinition}
D_{mn}(l)&= \int \frac{\mathrm{d}^2\mathbf{q}}{(2\pi)^2}
\frac{\mathrm{d}^2\mathbf{q'}}{(2\pi)^2}\; f_m(\sfrac{\mathbf{l}}{2}-
\mathbf{q}) f_n(\sfrac{\mathbf{l}}{2}-\mathbf{q'})
\Phi_{\mtin{SC}}(q,q',l)_{\big|q_0=q_0'=\frac{l_0}{2}} \, .
\end{align}
With this choice of definition, (RAS) implies
$D_{mn}(l)=D_{nm}(l)=\sigma_m\sigma_n D_{mn}(l)$ where
$f_n(\mathbf{p})=\sigma_n f_n(-\mathbf{p})$ with $\sigma_n$ either
$+$ or $-$. Therefore, $D$ is a block matrix with separate blocks
for singlet and triplet symmetry.

Likewise the magnetic and forward scattering channels are expanded.
Now by (RAS), $\Phi_{\mtin{M}/\mtin{K}}(q,q',l)=
\Phi_{\mtin{M}/\mtin{K}}(q',q,-l)$, so we can write
\begin{align}\label{eq:MKexpansion}
\Phi_{\mtin{M}}^{\Lambda}(q,q',l)&=\sum_{m,n\in\ca{I}_{\mtin{MK}}}
M_{mn}(l) f_m(\mathbf{q}+ \sfrac{\mathbf{l}}{2})
f_n(\mathbf{q'}-\sfrac{\mathbf{l}}{2}) + R_{\mtin{M}}(q,q',l)\\
\nonumber
\Phi_{\mtin{K}}^{\Lambda}(q,q',l)&=\sum_{m,n\in\ca{I}_{\mtin{MK}}}
K_{mn}(l) f_m(\mathbf{q}+ \sfrac{\mathbf{l}}{2})
f_n(\mathbf{q'}-\sfrac{\mathbf{l}}{2})+ R_{\mtin{K}}(q,q',l)
\end{align}
with symmetric boson propagators $M_{mn}(l)=M_{nm}(-l)$ and
$K_{mn}(l)=K_{nm}(-l)$ and remainder terms $R_{\mtin{M}}$ and
$R_{\mtin{K}}$. The functions $(f_n)_{n\in\ca{I}_{\mtin{MK}}}$ are
again scale--independent, orthonormalized on $(-\pi,\pi]^2$, and
frequency independent. Although different expansions for the
magnetic and forward scattering channel are possible, we choose the
same for notational simplicity. The index sets $\ca{I}_{\mtin{SC}}$
and $\ca{I}_{\mtin{MK}}$ need not be the same. 
In analogy to the superconducting boson propagator we set
\begin{align}\label{eq:MKdefinition}
M_{mn}(l)&= \int \frac{\mathrm{d}^2\mathbf{q}}{(2\pi)^2}
\frac{\mathrm{d}^2\mathbf{q'}}{(2\pi)^2}\; f_m(\mathbf{q}+
\sfrac{\mathbf{l}}{2}) f_n(\mathbf{q'}- \sfrac{\mathbf{l}}{2})
\Phi_{\mtin{M}}(q,q',l)_{\big|q_0=-\frac{l_0}{2} ,
q_0'=\frac{l_0}{2}}\\ \nonumber K_{mn}(l)&= \int
\frac{\mathrm{d}^2\mathbf{q}}{(2\pi)^2}
\frac{\mathrm{d}^2\mathbf{q'}}{(2\pi)^2}\; f_m(\mathbf{q}+
\sfrac{\mathbf{l}}{2}) f_n(\mathbf{q'}- \sfrac{\mathbf{l}}{2})
\Phi_{\mtin{K}}(q,q',l)_{\big|q_0=-\frac{l_0}{2} ,
q_0'=\frac{l_0}{2}} \,\, .
\end{align}
With the boson propagators defined in this way the three expansions
and the remainder terms fulfill (RAS) and (PHS) separately. The
magnetic interaction describes interacting spin operators. For
example, a constant $f_{\mtin{S}}(\mathbf{p})=1$ describes a local
spin operator and a further expansion in the $s$--wave channel
generalizes to nearest neighbors and next to nearest neighbors and
so on. A $d$--wave form factor in the forward scattering expansion
describes a possible Pomeranchuk instability \cite{HalbothMetznerPomeranchuk}.

So far we have decomposed the interaction vertex such that boson
propagators carry the critical momentum dependence of the right hand
side of the RG equation at least as long as the effective vertex
function is regular. If it can be shown that the remainder terms
$R_{\mtin{SC}}$, $R_{\mtin{M}}$, and $R_{\mtin{K}}$ remain regular
or at least less singular than the boson propagators (even if the
effective vertex function develops a singularity in the flow), then
several benefits are gained for the analysis of competing
instabilities. First, the decomposition of the interaction allows to
identify qualitatively which instabilities are favored. After the
flow is stopped, the dominant terms in the interaction can be
decoupled by Hubbard--Stratonovich transformations,
so that one can proceed with a bosonic flow. Furthermore, the parametrization
of the flow is simplified, since no function of three fermion momenta and frequencies has to be 
studied, but only several functions of one fermion momentum and frequency.
The boson fields involved have point singularities,
which pose substantially less numerical effort than the 
extended Fermi surface singularities.

\section{The Superconducting Channel\label{sec:superconductivity}}

In this section we consider an example to illustrate our method. Let
the Fermi surface be curved and regular.
By this we mean that the Fermi surface does not meet the Van Hove
points and that Umklapp scattering is irrelevant. Then the
particle--hole bubble is negligible compared to the
particle--particle bubble, which develops a strong peak for small
momentum and frequency.

In a first step the particle--hole graphs are altogether neglected.
The RG equation (\ref{eq:VertexFlow}) without selfenergy effects is
then solved by the solution of a self--consistency equation
\begin{align}\label{eq:SCselfconsistent}
V_{\Lambda}(q,l-q,q') = V_{\Lambda_0}(q,l-q,q') - \int \mathrm{d}
p\; G(p) G(l-p) V_{\Lambda_0}(q,l-q,p) V_{\Lambda}(l-p,p,q') \, ,
\end{align}
where $G(p)=[i p_0 - e(\mathbf{p})]^{-1} \chi_{\Lambda}(p)$ is the
free fermion propagator multiplied with a regulator function
$\chi_{\Lambda}(p)$. In this section we choose a strict Fermi
surface momentum cut--off, that is, modes with
$|e(\mathbf{p})|<\Lambda$ are suppressed completely. If the vertex
function is expanded in a complete set of $\Lambda$ independent
functions $f_n$ but $\Lambda$ dependent coefficients
$\nu^{\Lambda}_{mn}(l)$
\begin{align*}
V_{\Lambda}(q,l-q,q')=V_{\Lambda}(q',l-q',q)= \sum_{m,n}
\nu_{mn}^{\Lambda}(l) f_m(\sfrac{{l}}{2}-{q})
f_n(\sfrac{{l}}{2}-{q'})\, ,
\end{align*}
then the self--consistency equation (\ref{eq:SCselfconsistent}) is
solved by $\nu^{\Lambda}=A^{-1}$, where the scale--dependent matrix
$A_{mn}(l)= (\nu^{\Lambda_0})^{-1}_{mn}(l) +
\Phi_{\mtin{pp}}^{mn}(l)$ is given by the initial condition and the
particle--particle bubble
\begin{align}
\Phi_{\mtin{pp}}^{mn}(l)&= \int \mathrm{d}p \; G(p) G(l-p)
f_m(\sfrac{{l}}{2} -{p}) f_n(\sfrac{{l}}{2} -{p}) \, .
\end{align}
We concentrate on the singular case $l=0$. If the functions $f_n$
could be chosen such that $\Phi_{\mtin{pp}}^{mn}(0)$ and
$(\nu^{\Lambda_0})^{-1}_{mn}(0)$ are both diagonal, then the matrix
$A(0)$ could easily be inverted to give $\nu^{\Lambda}_{mn}(0) =
\nu^{\Lambda_0}_n(0) [1+\nu^{\Lambda_0}_n(0)
\Phi_{\mtin{pp}}^{nn}(0)]^{-1} \delta_{mn}$. If there is an
attractive channel in the initial condition, that is
$\nu_n^{\Lambda_0}<0$ for some $n$, the flow in this channel
diverges for large enough $\beta$ due to a zero in the denominator.
On the other hand, the flow of $\nu_n^{\Lambda}$ is asymptotically
free for all $n$ with $\nu_n^{\Lambda_0}>0$ (repelling).

However, $\Phi_{\mtin{pp}}^{mn}(0)$ is not diagonal in general.
Suppose that the functions $f_n$ do not depend on frequency.
Following \cite{SalmhoferBook} we change variables $\mathbf{p}=\pi(E,\theta)$
with $E=e(\mathbf{p})$ and an angle $\theta$ to obtain
\begin{align*}
\Phi_{\mtin{pp}}^{mn}(0) = \int\limits_{\Lambda}^{\Lambda_0}
\mathrm{d} E\; \frac{\tanh \frac{\beta E}{2}}{2E} \int
\mathrm{d}\theta\;
J(E,\theta) f_m(\pi(E,\theta))f_n(\pi(E,\theta))
\end{align*}
with Jacobian $J(E,\theta)$. The $\theta$--integral is a smooth
function of $E$ for a curved and regular Fermi surface. For
$\Lambda_0$ small enough the zeroth order of a Taylor expansion in
$E$, that is a projection onto the Fermi surface, gives diagonality
if the $f_n$'s are chosen as Fermi surface harmonics
\cite{Fermisurfaceharmonics}. Therefore the main contribution of
$\Phi_{\mtin{pp}}^{mn}(0)$ can be made diagonal for a curved and
regular Fermi surface.

This gives a good understanding of the flow in the superconducting
channel (\ref{eq:SCexpansion}). In a diagonal expansion of dominant
processes only form factors that give rise to positive boson
propagators have to be taken into account. All other terms in the
expansion are suppressed to zero and can be neglected since they
will not influence the qualitative behavior of the flow. In practice
one suspects that only the biggest positive coefficient plays a role
in the flow to strong coupling (depending on the size of the
corresponding bubble).

For a non--diagonal expansion one cannot separate the irrelevant
modes exactly. There will be a flow to strong coupling if $\det A=0$
(for any finite expansion) at some scale. However, since diagonality
holds approximately, there is a good chance to capture the singular
behavior of $A$ in a small matrix of well--chosen form factors.

In a second step we take into account the full one--loop RG equation
with all particle--particle and particle--hole graphs and start with
the initial repulsive Hubbard interaction. Due to the curved and regular Fermi
surface and negligible Umklapp scattering assumed in this section,
we only consider the superconducting channel and decompose
\begin{align}\label{eq:VSCdecomposition}
V_{\Lambda}(q,l-q,q')&= -\sum_{m,n\in\ca{I}_{\mtin{SC}}} D_{mn}(l)
f_m(\sfrac{\mathbf{l}}{2} -\mathbf{q})
f_n(\sfrac{\mathbf{l}}{2}-\mathbf{q'})  + U + R(q,l-q,q') \, .
\end{align}
As discussed above we select only a few form factors that describe
the superconducting channel correctly and drop the remainder term
$R_{\mtin{SC}}$ of the expansion in the superconducting channel
(\ref{eq:SCexpansion}).\footnote{In particular, since the form
factors do not depend on frequency, we assume that the frequency
dependence of the vertex function is described well by the singular
frequency $l_0$.} The function $R$ in (\ref{eq:VSCdecomposition})
arises from particle--hole graphs. We will study its influence on
the flow in the superconducting channel and show that the
Kohn--Luttinger effect is present in our method. That is, although
starting at $\Lambda_0$ with a repelling
$V_{\Lambda_0}(q,l-q,q')=U>0$, particle--hole terms will create an
attractive superconducting interaction.

Inserting the decomposition (\ref{eq:VSCdecomposition}) into the RG
equation (\ref{eq:VertexFlow}) and projecting the particle--particle
graphs according to (\ref{eq:Ddefinition}) gives the flow equation
for the superconducting boson propagators for $n,m\in\ca{I}_{\mtin{SC}}$
\begin{align}
\dot{D}_{mn}(l) &= \sigma_n \int \mathrm{d}\mu(p,l-p) \left[
\sum_{a\in\ca{I}_{\mtin{SC}}}
  D_{ma}(l) f_a(\sfrac{\mathbf{l}}{2} -\mathbf{p}) - \delta_{m,\mtin{S}} U -
    \alpha_m(p,l)\right]  \\  \nonumber
& \hspace{4cm}  \times\left[ \sum_{a'\in\ca{I}_{\mtin{SC}}}
  D_{na'}(l) f_{a'}(\sfrac{\mathbf{l}}{2} -\mathbf{p}) - \delta_{n,\mtin{S}} U -
    \alpha_n(p,l)\right]
\end{align}
with initial values $D_{mn}(l)=0$ at scale $\Lambda_0$ and with the
contribution from the particle--hole channels $\alpha_m(p,l) = \int
\frac{\mathrm{d}^2\mathbf{q}}{(2\pi)^2} f_m(\frac{\mathbf{l}}{2}
-\mathbf{q}) R(q,l-q,p)_{|q_0=l_0/2}$. We denote $\mathrm{d}
\mu(p,k)=\frac{\mathrm{d}}{\mathrm{d}\Lambda} \left(G(p)G(k)\right)\mathrm{d}p$
for the bubble integration and $f_m(-\mathbf{p})=\sigma_m
f(\mathbf{p})$ distinguishes between singlet and triplet symmetry.
Further we assume that $\ca{I}_{\mtin{SC}}$ contains the index
$n=\mtin{S}$ where $f_{\mtin{S}}(\mathbf{p})=1$ is the constant form
factor.

The evolution of $R$ is given by the particle--hole graphs, which
unlike the particle--particle graphs remain bounded at all scales. In
the beginning of the flow $D$ is still small, so we neglect all
particle--hole graphs proportional to $D$. We further neglect the
remaining direct particle--hole graphs since they cancel 
excactly for a constant vertex function, hence remain small for a
nearly constant vertex function. (The initial condition is
$V_{\Lambda_0}=U$, that is, constant.) Then the flow equation for
$R$, given by crossed particle--hole graphs only, is solved by
$R(q,l-q,q')=\psi(q'-q)-U$ with
$\psi(l)=U[1+U\Phi_{\mtin{ph}}(l)]^{-1}$ where
$\Phi_{\mtin{ph}}(l)=\int \mathrm{d}p\; G(p)G(p+l)$ is the
particle--hole bubble.

Although $\Phi_{\mtin{ph}}(l)<0$ (for $l_0=0$), the denominator of
$\psi$ remains nonzero for a sufficiently small initial coupling $U$
since the particle--hole bubble is bounded in case of a curved and
regular Fermi surface. So $\psi(l)$ is a regular (smooth for an
appropriate cut--off) and symmetric function that can be expanded in
the form factors we are interested in. Again, 
 the  frequency dependence of $\psi$ is not taken into account here. Neglecting possible
remainders of this expansion, the flow equation for the
superconducting boson propagator reads
\begin{align}\label{eq:KohnLutt} \dot{D}_{mn}(l)&= \sigma_n
\sum_{a,a'\in\ca{I}_{\mtin{SC}}} \Big[ D_{ma}(l)-b_{ma}\Big]
\dot{\Phi}_{\mtin{pp}}^{aa'}(l) \Big[D_{a'n}(l) - b_{a'n}\Big] \, ,
\end{align}
where $b_{aa'}=\int\frac{\mathrm{d}^2\mathbf{q}}{(2\pi)^2}
\frac{\mathrm{d}^2\mathbf{q'}}{(2\pi)^2} \,\psi(q'-q)_{\big|
q_0=q_0'} f_a(\mathbf{q}) f_{a'}(\mathbf{q'})$.

If we only take into account two form factors
$\ca{I}_{\mtin{SC}}=\{\mtin{S},1\}$ with $f_{\mtin{S}}(\mathbf{p})=1$
($s$--wave) and $f_1(\mathbf{p})=\cos p_x -\cos p_y$ ($d_{x^2-y^2}$--wave),
then $b_{\mtin{S}1}=b_{1\mtin{S}}=0$, so b is diagonal as is
$\Phi_{\mtin{pp}}^{aa'}(0)$ in this case. Then the boson propagator
$D_{mn}(0)$ is diagonal as well and the flow equation
(\ref{eq:KohnLutt}) factorizes for $l=0$ such that the right hand
side is always positive. The coefficient $b_{\mtin{S}\mtin{S}}$ is
of order $U$ and positive, so $D_{\mtin{S}\mtin{S}}$ is suppressed
to zero. That is, $s$--wave superconductivity is always suppressed
by the initial repulsive interaction. On the other hand $b_{11}$ is
of order $U^2$ and negative, so $D_{11}(0)$ grows without bounds. 
Therefore, particle--hole fluctuations induce an attractive $d$--wave
pairing interaction, which will be essential for the study of
superconductivity in the Hubbard model at Van Hove filling. Here we
have not included triplet superconductivity, which can become dominant as well.
In general, attractive interactions in the Cooper channel are induced, so 
our method does capture the Kohn--Luttinger effect. 
The approximations made are valid provided the scale is not too low.

In summary, the proposed decomposition of the interaction vertex is
well understood for a curved and regular Fermi surface, where there
is no Van Hove singularity and where Umklapp scattering is
irrelevant. Namely, if the expansion in form factors is diagonal, we
gave a clear argument which form factors have to be included in the
expansion, such that the remainder of the expansion is negligible.
In the general case diagonality holds approximately. The influence
of subdominant particle--hole fluctuations can become important if
there is no strong attractive interaction already initially.

However, if Umklapp scattering and a Van Hove singularity in the
density of states are present, particle--hole fluctuations can
diverge by themselves and the mixing between particle--particle and
particle--hole channels is strong. This case is not analytically
understood yet and treated numerically in the following.

\section{$\Omega$--Scheme and Integration over High Scales \label{sec:regularization}}

Although the renormalization group idea is independent
of the type of regularization used to define the flow 
(provided the regularization satisfies some minimal conditions,
e.g.\ that it makes the flow equation well--defined),
its choice can become a delicate matter 
once approximations are used.
In particular, we restrict the flow to the symmetric phase
and apply the one--loop truncation. Parts of the vertex function
will grow strongly at small scales, indicating an instability to an
ordered (symmetry broken) state. Since we are using a weak coupling
scheme, we have to stop the flow before all modes are integrated
out, that is, the full model is not recovered in the flow.
Therefore, the one--loop flow confined to the symmetric phase
may depend on the type of regularization, 
and one has to compare the results of different 
regularization schemes.

Since we are especially interested in the interplay between
$d$--wave superconductivity and ferromagnetism,  we have to choose a regularization that does not suppress small momentum particle--hole fluctuations \cite{TemperatureFlow}.
In the Hubbard model at Van Hove filling the density of states
diverges logarithmically at low single--particle energies. At zero
temperature and without scale regularization, the particle--hole bubble at
zero momentum transfer is given by minus the density of states.
Therefore, the particle--hole bubble at zero momentum transfer and
zero temperature should diverge logarithmically at low energy scales in the presence of a Van Hove singularity. However, for example, this is not the case for a Fermi
surface cut--off which, at all nonzero scales, 
suppresses the small--momentum particle--hole excitations
completely (see the discussion in \cite{TemperatureFlow}).
For such a Fermi surface cut--off the particle--hole bubble at zero
momentum transfer is zero for all finite scales at temperature zero.
Even for nonzero temperature this means an artificial suppression of
small momentum particle--hole processes down to
the very lowest scales.
Instead of a Fermi surface cut--off we multiply the bare propagator
with the regulator function
\begin{align}\label{eq:OmegaCutoff}
\chi_{\Omega}(p)=\frac{p_0^2}{p_0^2+\Omega^2} \, ,
\end{align}
which is independent of momentum but depends on frequency. The
frequency $\Omega$ is the scale parameter we use to
generate the RG flow and it replaces the $\Lambda$ we used in the general discussion in Section \ref{sec:RenormalizationGroup}. Since
possible zeros of the denominator in the integrand of the bubble
integration are canceled, the bubbles are regularized for all
temperatures if the scale $\Omega>0$. Small momentum particle--hole
processes are not artificially  suppressed to lowest scales with
this regularization since the particle--hole bubble, now formed with
the $\Omega$--dependent propagator $G(p)=\big[ip_0 - e(\mathbf{p})\big]^{-1}
\chi_{\Omega}(p)$,
 has the right asymptotic
scaling behavior at temperature zero and Van Hove filling, that is
$\Phi_{\mtin{ph}}(0) \sim \log \Omega$. The particle--particle
bubble $\Phi_{\mtin{pp}}(0)\sim (\log \Omega)^2$ diverges faster for the same parameters, as expected.

In the limit $\Omega\to\infty$ the regulator function
(\ref{eq:OmegaCutoff}) vanishes, hence the vertex function is the
initial on--site interaction $U$ of the Hubbard model.
For $\Omega \to 0$, $\chi_\Omega (p) \to 1$, 
i.e.\ the regulator is removed. The regularization is mild, 
but it suffices to make all loop integrals in the flow equation
converge. Thus it is an admissible scheme. For brevity we call this regularization the $\Omega$--scheme.

Because perturbation theory converges for large $\Omega$ \cite{PositivityConvergence,MatsubaraUV}, we fix a scale $\Omega_0$ 
and perform the integration over the degrees of freedom 
with propagator $(1 - \chi_{\Omega_0}(p))\big[ip_0 - e(\mathbf{p})\big]^{-1}$ by perturbation theory. The result provides the initial condition for the flow 
at scale $\Omega=\Omega_0$.
To second order in $U$ this gives for the Hubbard model 
in terms of a vertex function as in equation (\ref{eq:vertexfunction}), \begin{align}\label{eq:vertexperturbation}
V_{\Omega_0}(p_1,p_2,p_3)=U - U^2\Phi_{\mtin{pp}}(p_1+p_2) - U^2
\Phi_{\mtin{ph}}(p_3-p_1) + \ca{O}(U/\Omega_0)^3 \, ,
\end{align}
where the first term is the initial Hubbard repulsion. The second
term is the particle--particle bubble, which generates an attractive
$s$--wave superconducting channel. It will always be dominated by
the repulsion $U$. The third term in (\ref{eq:vertexperturbation})
comes from the crossed particle--hole graph. The
contribution from the direct particle--hole graphs 
cancels out because the Hubbard interaction is on--site.

We choose the initial condition for the boson propagators at scale
$\Omega_0$ as
\begin{align}\label{eq:initialbosonprop}
D_{\mtin{S}\mtin{S},0}(l) &= U^2\Phi_{\mtin{pp}}(l)  \\ \nonumber
M_{\mtin{S}\mtin{S},0}(l)&=K_{\mtin{S}\mtin{S},0}(l)
=-U^2\Phi_{\mtin{ph}}(l)>0
\end{align}
for the constant $s$--wave form factor $f_{\mtin{S}}(\mathbf{p})=1$.
The symmetric part of $D_{\mtin{S}\mtin{S},0}(l)$ is positive as well,
as is $D_{\mtin{S}\mtin{S},0}(l_0=0,\mathbf{l})$.
This assignment of the result of perturbation theory to the initial
boson propagators is not unique. For example, attracting parts of the
particle--hole term could be absorbed in a superconducting
$d$--wave term. But the assignment (\ref{eq:initialbosonprop})
is consistent with the definition of the three channels in Section
\ref{sec:decomposition}, which do not mix in second order.
Furthermore, the Kohn--Luttinger effect is very small in
perturbation theory. The attractive coefficient of
$d_{x^2-y^2}$--superconductivity in an expansion of
$U^2\Phi_{\mtin{ph}}(p_3-p_1)$ is bounded by $\sim 0.01U^2$ even for low scales and remains finite if the scale regularization is removed. Therefore the ambiguity of (\ref{eq:initialbosonprop})
will not be important. In fact, we have numerically checked
different distributions of the interaction terms in the initial conditions. 
We choose $\Omega_0>U$ big enough such that the results presented in Section \ref{sec:results} do not depend on the initial conditions for the boson propagators. The latter are negligible compared to interactions generated in the flow by the local repulsion term linear in $U$. So for large enough $\Omega_0$ our results are independent of the particular choice of $\Omega_0$.

\section{RG Equations of the Boson Propagator
  Flow\label{sec:RGequations}}

In order to derive the RG equations for the boson propagators we
insert the expansion of the three channels (\ref{eq:SCexpansion})
and (\ref{eq:MKexpansion}) into the RG equation
(\ref{eq:VertexFlow}) and project the right hand side of it
according to the definitions of the boson propagators
(\ref{eq:Ddefinition}) and (\ref{eq:MKdefinition}). Similarly,
equations for the remainder terms $R_{\mtin{SC}}$, $R_{\mtin{M}}$,
and $R_{\mtin{K}}$ can be obtained.

In case of a curved and regular Fermi surface
we discussed in Section \ref{sec:superconductivity} how to separate
singular from regular parts in the superconducting channel if the
other channels are neglected. That gave a condition which form
factors should be included in the expansion of the superconducting
channel. Furthermore, as also analyzed in Section
\ref{sec:superconductivity}, we developed a good
understanding of the integration of all one--loop graphs down to
intermediate scales in this case. Therefore we argue that, with the
right choice of form factors, the remainder terms can be dropped
because the channels are described well even by only a few boson
propagators.

In presence of Van Hove singularities and relevant Umklapp
scattering, however, the mixing of the channels is very hard to
control. Here we rely on the results of
\cite{ZanchiSchulz2000,HalbothMetzner,HalbothMetznerPomeranchuk,NpatchRG,Umklapp} and choose a combination of form
factors corresponding to the main instabilities found there.
A detailed study of the remainder terms is left for future work.

\begin{figure} [h]
\centering
\includegraphics[width=.8\textwidth]{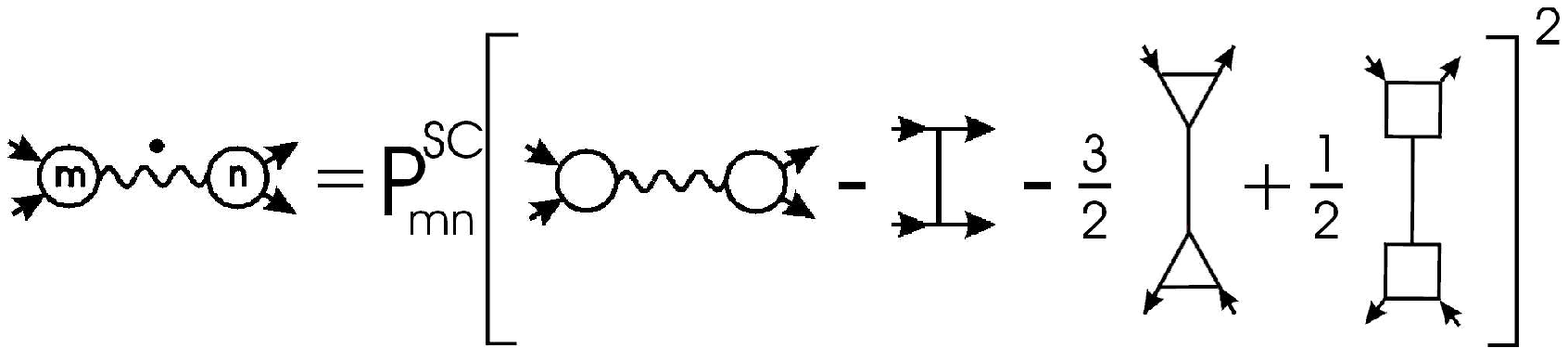}
\caption{The graphical representation of the flow in the
  singlet superconducting channel.\label{fig:GraphSC} }
\end{figure}

The flow equations derived in this section are valid for arbitrary
index sets $\ca{I}_{\mtin{SC}}$ and $\ca{I}_{\mtin{MK}}$, assuming
they both contain the index $\mtin{S}$ that stands for the constant
form factor $f_{\mtin{S}}(\mathbf{p})=1$. If the remainder terms are
dropped, then the flow equation for the superconducting boson
propagator reads for $n,m\in\ca{I}_{\mtin{SC}}$
\begin{align}\label{eq:Dflow}
\dot{D}_{mn}(l) &= \sigma_m \int \mathrm{d}\mu(p,l-p)
\Big[\sum_{a\in\ca{I}_{\mtin{SC}}} D_{ma}(l)
f_a(\sfrac{\mathbf{l}}{2}-\mathbf{p}) -U\delta_{m\mtin{S}}
-\alpha_m^{\mtin{SC}}(p,l)\Big]
\\\nonumber &\hspace{4cm} \times\Big[\sum_{a\in\ca{I}_{\mtin{SC}}}
D_{an}(l) f_a(\sfrac{\mathbf{l}}{2}-\mathbf{p}) -U\delta_{n\mtin{S}}
-\alpha_n^{\mtin{SC}}(p,l)\Big]\, ,
\end{align}
where $\mathrm{d}\mu(p,l\pm p)=\mathrm{d}p
\frac{\mathrm{d}}{\mathrm{d}\Omega} G(p)G(l\pm p)$ is the bubble
integration. The  initial local repulsion $U$ will suppress
$s$--wave superconductivity $D_{\mtin{S}\mtin{S}}$.
The contribution of the magnetic and forward scattering channels is
given by
\begin{align}\nonumber
\alpha_m^{\mtin{SC}}(p,l)&= \frac 12 \int
\frac{\mathrm{d}^2\mathbf{q}}{(2\pi)^2} \;
f_m(\sfrac{\mathbf{l}}{2}-\mathbf{q})
\sum_{b,b'\in\ca{I}_{\mtin{MK}}}
f_b(\sfrac{\mathbf{p}+\mathbf{q}}{2})f_{b'}(\mathbf{l}-
\sfrac{\mathbf{p}+\mathbf{q}}{2})\\ &\nonumber
\hspace{3cm}\times\Big[ (2+\sigma_m) M_{bb'}(p-q) - \sigma_m
K_{bb'}(p-q)\Big]_{\Big|q_0=\frac{l_0}{2}}\,\, .
\end{align}
For example, the magnetic boson propagator $M_{\mtin{S}\mtin{S}}$
induces an attracting interaction for $d_{x^2-y^2}$--wave
superconductivity as described in Section
\ref{sec:superconductivity}. The corresponding diagrams of the flow
equation (\ref{eq:Dflow}) are plotted in Figure \ref{fig:GraphSC}
for singlet symmetry $\sigma_m=+1$ in a symbolic way.
The graphs in the brackets are the superconducting, initial Hubbard
repulsion, magnetic, and forward scattering interaction
respectively. The square on the right hand side means 
that any two of the graphs in the bracket are put beside 
one another and connected to form one--loop diagrams.
This generates three types of graphs, namely 
direct, vertex correction, and box graphs (see Figure 2). 
The latter two are not of the form of a boson propagator 
mediating between two fermion bilinears. 
We expand them in a sum of terms of that form 
(given by our ansatz for the interaction)
and drop the remainder term. 
To get the coefficients in the expansion (\ref{eq:SCexpansion}),
we apply the projections equation. 

\begin{figure} [h]
\centering
\includegraphics[width=.45\textwidth]{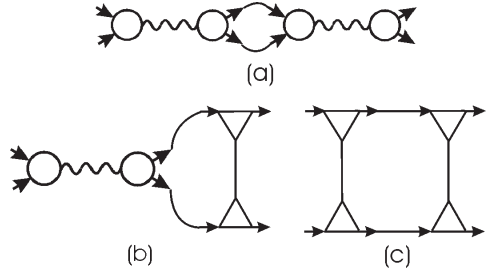}
\caption{Three examples of arising graphs if the square in Figure
\ref{fig:GraphSC} is taken: (a) the direct graph, (b) a vertex
correction graph, and (c) a box graph.\label{fig:GraphSCexamples} }
\end{figure}

In order to illustrate how the square in Figure \ref{fig:GraphSC} is
applied, we state two examples. Taking two superconducting
interactions and connecting them with two fermion propagators such
that they form a loop, gives the direct graph in Figure
\ref{fig:GraphSCexamples}(a). If the loop integration is performed,
it is again of the form of two Cooper pairs interacting via a boson.
By connecting a superconducting interaction with a magnetic
interaction a vertex correction graph arises, given in Figure
\ref{fig:GraphSCexamples}(b). This is not of the demanded form, so
the projection is nontrivial. A box diagram, for example, is formed
by connecting two magnetic interactions, see Figure \ref{fig:GraphSCexamples}(c).

\begin{figure} [h]
\centering
\includegraphics[width=.8\textwidth]{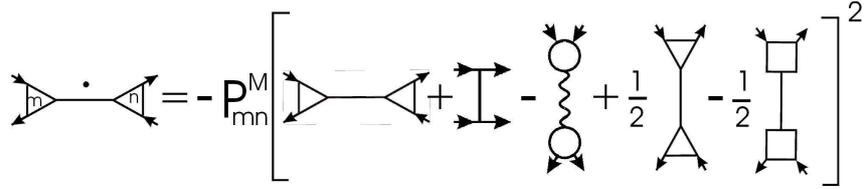}
\caption{The flow in the
  magnetic channel.\label{fig:GraphM} }
\end{figure}

In the magnetic channel the flow is described by
\begin{align}\label{eq:MagneticFlow}
\dot{M}_{mn}(l) &= - \int \mathrm{d}\mu(p,p+l)
\Big[\sum_{b\in\ca{I}_{\mtin{MK}}} M_{mb}(l)
f_b(\mathbf{p}+\sfrac{\mathbf{l}}{2}) +U\delta_{m\mtin{S}}
+\alpha_m^{\mtin{M}}(p,l)\Big] \\ \nonumber &\hspace{4cm}
\times\Big[\sum_{b\in\ca{I}_{\mtin{MK}}} M_{bn}(l)
f_b(\mathbf{p}+\sfrac{\mathbf{l}}{2})
+U\delta_{n\mtin{S}}+\alpha_n^{\mtin{M}}(p,l)\Big]\, ,
\end{align}
where
\begin{align*}
\alpha_m^{\mtin{M}}(p,l)&= \frac 12
\int\frac{\mathrm{d}^2\mathbf{q}}{(2\pi)^2}   \;
f_m(\mathbf{q}+\sfrac{\mathbf{l}}{2}) \left[
  -2\sigma_m\sum_{a,a'\in\ca{I}_{\mtin{SC}}}
D_{aa'}(p-q) f_a(\sfrac{\mathbf{p}+\mathbf{q}}{2}+\mathbf{l})
f_{a'}(\sfrac{\mathbf{p}+\mathbf{q}}{2})\right. \\ \nonumber
&\left.\hspace{0.5cm} + \sum_{b,b'\in\ca{I}_{\mtin{MK}}}
f_b(\sfrac{\mathbf{p}+\mathbf{q}}{2})f_{b'}(\mathbf{l}+
\sfrac{\mathbf{p}+\mathbf{q}}{2})\Big[M_{bb'}(p-q)
-K_{bb'}(p-q)\Big]\right]_{\Big|q_0=-\frac{l_0}{2}}\,\, .
\end{align*}
Again the initial Hubbard interaction contributes only to 
the local $s$--wave coupling part. It will drive the boson propagator
$M_{\mtin{S}\mtin{S}}$. In contrast to that, the superconducting
$s$--wave boson propagator $D_{\mtin{S}\mtin{S}}$ will screen the
local magnetic interaction $M_{\mtin{S}\mtin{S}}$. Because this
effect is quite substantial, especially if $U$ is not very small,
$D_{\mtin{S}\mtin{S}}$ is important and must be included 
in the flow, even though
it does not become singular. The influence of $d_{x^2-y^2}$
superconductivity with form factor $f_1(\mathbf{p})=\cos p_x - \cos p_y$ depends
strongly on the boson momentum $\mathbf{l}$. It will suppress
$M_{\mtin{S}\mtin{S}}(l)$ near $\mathbf{l}=0$, that is,
suppress ferromagnetism, 
and enhance $M_{\mtin{S}\mtin{S}}(l)$ near
$\mathbf{l}=\hat{\pi}=(\pi,\pi)$, that is, antiferromagnetism, since
$f_1(\mathbf{p}+\hat{\pi})=-f_1(\mathbf{p})$. The symbolic
representation with graphs of the flow equation
(\ref{eq:MagneticFlow}) is given in Figure \ref{fig:GraphM}.

\begin{figure} [h]
\centering
\includegraphics[width=.8\textwidth]{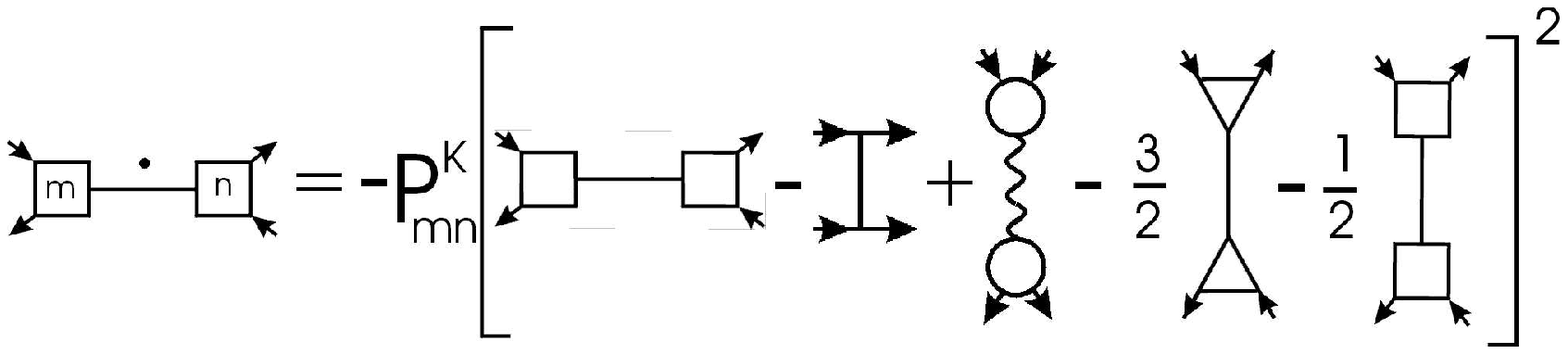}
\caption{The flow in the
  forward scattering channel.\label{fig:GraphK} }
\end{figure}

Finally the flow equation for the forward scattering channel is
given by
\begin{align}
\dot{K}_{mn}(l) &= - \int \mathrm{d}\mu(p,p+l)
\Big[\sum_{b\in\ca{I}_{\mtin{MK}}} K_{mb}(l)
f_b(\mathbf{p}+\sfrac{\mathbf{l}}{2}) -U\delta_{m\mtin{S}}
-\alpha_m^{\mtin{K}}(p,l)\Big] \\ \nonumber &\hspace{5cm}
\times\Big[\sum_{b\in\ca{I}_{\mtin{MK}}} K_{bn}(l)
f_b(\mathbf{p}+\sfrac{\mathbf{l}}{2}) - U\delta_{m\mtin{S}} -
\alpha_n^{\mtin{K}}(p,l)\Big]
\end{align}
with
\begin{align*} \alpha_m^{\mtin{K}}(p,l)&= \frac 12
\int\frac{\mathrm{d}^2\mathbf{q}}{(2\pi)^2}  \;
f_m(\mathbf{q}+\sfrac{\mathbf{l}}{2}) \left[2\sigma_m
\sum_{a,a'\in\ca{I}_{\mtin{SC}}} (1-2\sigma_a) D_{aa'}(p-q)
f_a(\sfrac{\mathbf{p}+\mathbf{q}}{2})\right.\\ &\hspace{8cm} \times
\;
f_{a'}(\mathbf{l}+\sfrac{\mathbf{p}+\mathbf{q}}{2}) \\
\nonumber &\left.\hspace{0.5cm} + \sum_{b,b'\in\ca{I}_{\mtin{MK}}}
f_b(\sfrac{\mathbf{p}+\mathbf{q}}{2})
f_{b'}(\mathbf{l}+\sfrac{\mathbf{p}+\mathbf{q}}{2})
\Big[3M_{bb'}(p-q) +K_{bb'}(p-q)\Big]\right]_{\Big|q_0=-\frac{l_0}{2}}\, \, .
\end{align*}
Similarly to $s$--wave superconductivity $K_{\mtin{S}\mtin{S}}$ is
suppressed by $U$. In analogy to the Kohn--Luttinger effect a
Pomeranchuk boson propagator $K_{11}$ is induced by magnetic
correlations. Graphically the equation is sketched in Figure
\ref{fig:GraphK}.

We finish this section with some remarks on the just derived RG
equations. 

1. Instead of the Hubbard model consider a mean--field
model. So set $U=0$ and let the initial condition for the boson
propagators be given by $D_{mn,0}(l)=D_{mn}^0 \delta_{l,0}$ and
similarly for $M$ and $K$. Then the flow equations recover
mean--field theory in the thermodynamic limit. Only the direct
graphs contribute in the thermodynamic limit since the $\alpha$'s
acquire an additional $\frac{1}{\beta L^2}$ factor.

2. Observing the flow equations, we find that the diagonal parts of the
boson propagators at frequency zero are monotonically increasing.
This is no contradiction to previous $N$--patch schemes where the
measured susceptibilities did not need to be monotone. For example,
the one--particle irreducible part of the connected expectation
value
\begin{align*}
-\left\langle \int
\mathrm{d}p\,f_m(\frac{\mathbf{l}}{2}-\mathbf{p})\ol{\psi}(q)
\frac{\epsilon}{2} \ol{\psi}(l-q) ; \int
\mathrm{d}p\,f_n(\frac{\mathbf{l}}{2}-\mathbf{p})\psi(q)
\frac{\epsilon}{2} \psi(l-q)\right\rangle \, ,
\end{align*}
which corresponds to the singlet superconducting susceptibility with
$f_n(p)=f_n(-p)$, is here given by $D_{mn}(l)$ plus a contribution
from the other channels. If the latter prevail, the slope of the
subdominant susceptibility can change sign even for $n=m$ and $l=0$.

3. The flow equations stated so far are quite general. They are valid
for arbitrary sets of form factors $\ca{I}_{\mtin{SC}}$ and
$\ca{I}_{\mtin{MK}}$. Also, no assumptions have been made concerning
the frequency and momentum dependence of the boson propagators. For
a numerical solution we specify this momentum dependence in the next
section. For given $\ca{I}_{\mtin{SC}}$ and $\ca{I}_{\mtin{MK}}$
this leads to substantial simplifications.

\section{Momentum Dependence of the Boson
  Propagators\label{sec:patching}}

While we have chosen particular functions for the form factors, the
frequency and momentum dependence of the boson propagators is not
specified yet. If this dependence is not constrained, the RG
equation for the boson propagators still remains a system of integro
differential equations which is hard to solve. For the
following numerical calculation we neglect the frequency dependence
of the boson propagators. Together with the frequency independent
form factors, this corresponds to a vertex function that does not
depend on frequency at all. The right hand side of the flow
equations is evaluated at zero boson frequency, which gives the main
contribution.

Furthermore we approximate the boson propagators by two--dimensional
step functions. That is, their momentum dependence is discretized by
dividing momentum space into segments on which the boson propagators
are constant. Then the RG equation is equivalent to a system of $x$
ordinary differential equations where $x$ is the number of boson
propagators times the number of segments. 
Note that the segments chosen here are neighborhoods of a
single point because the boson propagators have point 
singularities. So they are different from the patches around
the Fermi surface used in $N$--patch studies. 

Thus the computational
effort is reduced compared to Fermi surface $N$--patch schemes,
where the frequency dependence is also neglected and $x\sim N^3$
with $N$ the number of patches for one 
fermion momentum. In many $N$--patch
studies $N$ is reduced by projecting the momentum dependence onto
the Fermi surface. This is not done here, we allow a general
discrete momentum dependence of the boson propagators. 
It will turn out in the RG flow that the singular momentum dependence of the boson propagators is mainly determined by the fermion one--loop bubbles. Therefore, contrary to $N$--patch schemes, we can choose the size and alignment of the segments to gain a more accurate approximation by step functions.

For small scales and low temperature the bubbles can 
develop strong
peaks at transfer momentum $\mathbf{l}=0$ and
$\mathbf{l}=\hat{\pi}=(\pi,\pi)$ (or at
$\mathbf{l}=\hat{\pi}-\delta$ for incommensurate Umklapp
scattering). Possible singularities of the boson propagators are
closely related to the corresponding direct graphs, which have the
same momentum dependence as the bubbles. Therefore we separate the
momentum dependence of the boson propagators into two parts
\begin{align}
B(\mathbf{l})= B^{(0)}(\mathbf{l}) \mathbbm{1}\left(|l_x|+|l_y|\le
\pi\right) + B^{(\hat{\pi})}(\mathbf{l}-\hat{\pi})
\mathbbm{1}\left(|l_x-\pi|+|l_y-\pi|\le \pi\right)
\end{align}
for $B=D$, $M$, or $K$ representing the different boson propagators
and $l_x,l_y\in [0,2\pi)$ and periodically continued elsewhere. We now approximate
$B^{(a)}(\mathbf{l})$, where $a=0$ or $\hat{\pi}$, by step functions for
$|l_x|+|l_y|\le \pi$ with $l_x,l_y\in(-\pi,\pi]$. Singularities will
appear for small $\mathbf{l}=(l_x,l_y)$ only. However, since we do
not choose $U \ll 1$, the magnetic local propagators
$M^{(a)}_{\mtin{S}\mtin{S}}(\mathbf{l})$ driven by $U$ are not
negligible for $\mathbf{l}$ away from zero. Likewise, the
propagators $D^{(a)}_{\mtin{S}\mtin{S}}(\mathbf{l})$ and
$K^{(a)}_{\mtin{S}\mtin{S}}(\mathbf{l})$ are of order $U$ for all
momenta. Therefore the step functions have to cover full momentum space and not just a small neighbourhood of $\mathbf{l}=0$.

\begin{figure} [h]
\centering
\includegraphics[width=.5\textwidth]{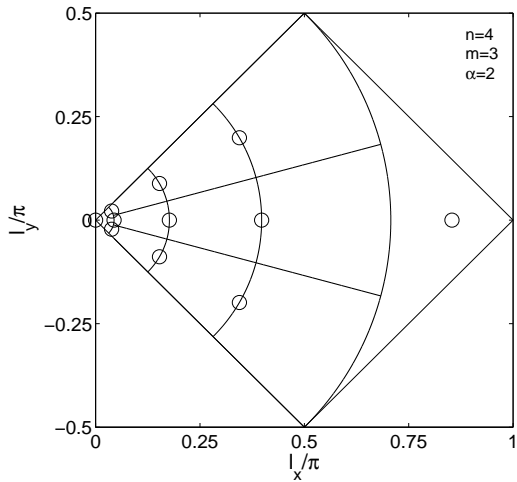}
\caption{Schematic plot of the segment sizes for $n=4$, $m=3$, and
$\alpha=2$. The circles mark the spots where the boson propagators
are evaluated. The momentum dependence of the boson propagators on
full momentum space is obtained by using
symmetries.\label{fig:patching}}
\end{figure}

Symmetries of the boson propagators can be used to reduce the number of step functions necessary.
If the boson propagators are diagonal, then they
obey the symmetries
$B^{(a)}(l_x,l_y)=B^{(a)}(-l_x,-l_y)=B^{(a)}(l_y,l_x)$ where $a=0$
or $\hat{\pi}$. In fact, for these symmetries to hold, the boson
propagators are allowed to consist of block matrices for special
form factors, such as $f_1$ and $f_3$. Changing to polar variables
$l_x=\rho\cos \phi$ and $l_y=\rho\sin \phi$, we divide the radial
coordinate $\rho\in[0,\frac{\pi}{\sqrt{2}}]$ into $n$ intervals as can
be seen in Figure \ref{fig:patching}. We choose smaller segments for small $\rho$, that is, for $k=1,\ldots n$ the segments 
are $\left[\left(\frac{k-1}{n}\right)^\alpha \frac{\pi}{\sqrt{2}} ,
\left(\frac k n \right)^\alpha \frac{\pi}{\sqrt{2}}\right]$, where
$\alpha\ge1$ is a parameter for choosing the radial size of the smallest
segment. The boson propagators are evaluated at the left border of
each interval. Using the symmetries stated above, the angular
variable $\phi$ needs only be discretized in
$\left[-\frac{\pi}{4},\frac{\pi}{4}\right]$. This range is divided
into $m$ homogeneous intervals, where the boson propagators are
evaluated at the midpoints. The $k=1$ radial segment has no angular
dependence imposed. Since we want to cover $|l_x|+|l_y|\le \pi$, one
extra segment is needed for the difference of the disk to the square.
This more inaccurate approximation for large momenta is justified since
the singularity  of the bubbles develops at small
momenta near $0$ or $\hat{\pi}$.

In the numerical calculation presented in the next section, we find
that the magnetic and forward scattering boson propagators have
sizeable contributions on a larger radial momentum range than the
superconducting boson propagators, which have a strong peak for very
small momenta only. This can also be observed in a comparison of the
particle--particle and the particle--hole bubbles. Therefore we
allow different $\alpha$'s for these two cases.

Omitting the frequency dependence and parameterizing the momentum
dependence with step functions simplifies the flow equations drastically
for given form factors. The Matsubara frequency sum over the loop
frequency $p_0$ can then be calculated explicitly. Also, by
trigonometric identities the $\mathbf{q}$--integrations
in the projections in $\alpha_{m}^{\mtin{SC}}$, $\alpha_{m}^{\mtin{M}}$, and
$\alpha_{m}^{\mtin{K}}$ become sums of terms, whose dependence on
the loop momentum $\mathbf{p}$ can be stated analytically. 
Thus only the momentum loop integral has to be performed numerically to
compute the flow.

\section{Numerical Results at Van Hove Filling\label{sec:results}}

We have numerically solved the flow equations for the boson
propagators with different sets of form factors. Already a flow with
only a constant form factor $f_{\mtin{s}}(\mathbf{p})=1$ in all
channels plus standard $d_{x^2-y^2}$--superconductivity with form factor
$f_1(\mathbf{p})=\cos p_x -\cos p_y$ in the superconducting channel
gives reasonable results. We have checked that including the higher $d_{x^2-y^2}$
harmonic $f_3(\mathbf{p})=\cos3p_x -\cos 3p_y$ in the
superconducting expansion produces only very small changes.
Here we present results that are obtained by expanding all channels
in the same set of form factors
$\ca{I}_{\mtin{SC}}=\ca{I}_{\mtin{MK}}=\{\mtin{S},1\}$. Since
$\Phi_{\mtin{pp}}^{1\mtin{S}}(a)=\Phi_{\mtin{ph}}^{1\mtin{S}}(a)=0$
for $a=0$ or $\hat{\pi}$, the boson propagators are diagonal at
momentum $\mathbf{l}=0$ or $\hat{\pi}$. Furthermore they are
diagonal for all $|l_x|=|l_y|$ and this does not change in the flow.
However, away from $|l_x|=|l_y|$ off--diagonal terms appear, which
at first are neglected here since the singular structure of the
boson propagators lies on these lines. At the end of this section we
present results where off--diagonal boson propagators are included.

The density is set to Van Hove filling, that is, when varying the
next to nearest neighbor hopping $-t'$ the particle number is
changed to fit $\mu=4t'$. With this choice of parameters the Hubbard model at weak coupling
has a strong tendency towards ferromagnetism.
Since we want to compare our results with those of the temperature RG flow
\cite{TemperatureFlow}, where ferromagnetism was first found in a
one--loop RG method for the Hubbard model, we choose $U=3$ as
well.\footnote{As noted before, all quantities are measured in units
of $t=1$.} Smaller couplings are harder to treat numerically.

While in principle our method can treat all temperatures, for
simplicity we here concentrate on temperature zero, which is not
accessible within the temperature RG flow. However, starting at
$\Omega_0=15$, we find  
that for all choices of $t'$, at least one of the boson propagators 
becomes almost singular at a point. In an approximation 
$(\mathbf{p}^2 + m_B^2)^{-1}$ for the boson propagator,  
one would say that the mass $m_B$ of the boson 
tends to zero. However, the propagators are not well--approximated
by this simple form, except maybe at very small momentum (depending
on the scale $\Omega$), 
and this is the main reason why we use a numerical method to capture
the momentum dependence. 
The flow to strong coupling observed in previous fermionic RG studies
is related to this pointlike singularity, 
in that certain approximations correspond to setting $p =0$, 
which hides the fact that the interaction only becomes strong
at points. 

\begin{figure} [h]
\centering
\includegraphics[width=.65\textwidth]{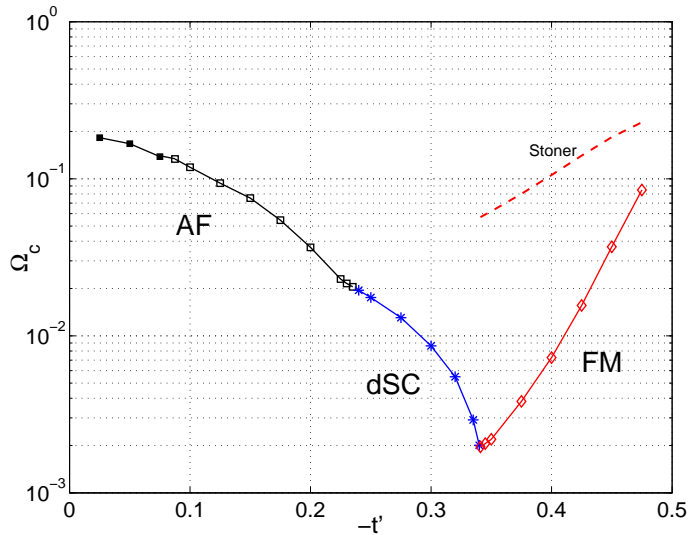}
\caption{The critical scale in dependence on hopping $-t'$ at
temperature zero and Van Hove filling for initial coupling $U=3$.
The instabilities of the Landau Fermi liquid are determined as
antiferromagnetism (AF), $d_{x^2-y^2}$ superconductivity (dSC), and
ferromagnetism (FM). The figure is further explained in the
text.\label{fig:OmegaKrit}}
\end{figure}

Although in our study, there is no flow to strong coupling in that sense, 
a true singularity in the interaction function introduces a significant 
change in power counting, which seems best to be captured by stopping
the fermionic flow at a certain scale $\Omega_{\mtin{C}}$,
performing a transformation to exchange bosons, 
and then attempting to continue to the symmetry--broken phase.
In our present study, we only consider the flow to the scale $\Omega_{\mtin{C}}$, which we call the critical scale,
and leave the other steps to future work. 
We stop the flow before the
maximum of any boson propagator reaches $B_{\max}=20$. The thus defined critical scale, which roughly corresponds to a critical temperature in the temperature flow, is plotted in Figure \ref{fig:OmegaKrit} over
different hopping $-t'$. The flow to strong coupling of a boson
propagator is interpreted as an instability to a corresponding
ordered state. However, this can only give a qualitative picture of
the actual ground state phase diagram. For the latter the flow has
to be continued in the symmetry broken phase. So for example, the
specific choice of $B_{\max}$ is arbitrary. However, it does not
effect the qualitative features of Figure
 \ref{fig:OmegaKrit}.
For $B_{\max}=16$, for example, the critical scale is at most about
15\% higher in the
magnetic regions and even less in the superconducting region.
Also, the range of $-t'$ where superconductivity dominates shrinks
by less than 1\%.

\begin{figure} [h]
\centering
\includegraphics[width=.68\textwidth]{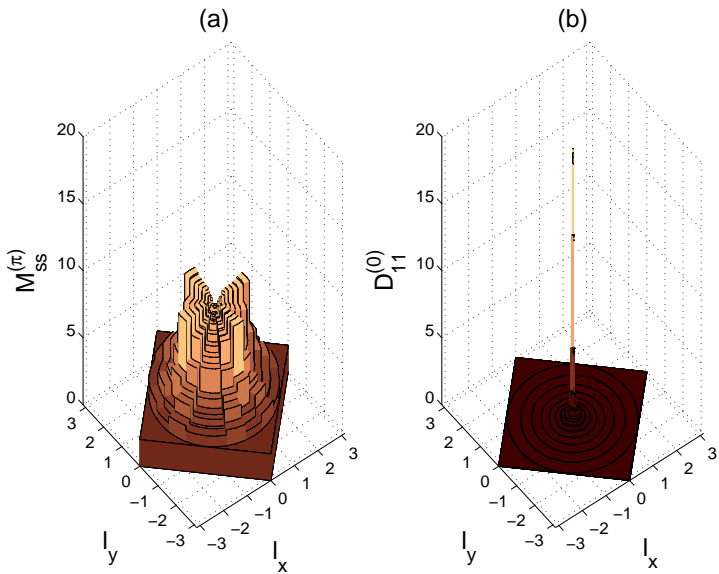}
\caption{Momentum dependence of the boson propagators (a)
$M_{\mtin{SS}}^{(\hat{\pi})}$ (antiferromagnetism) and (b)
$D_{11}^{(0)}$ ($d_{x^2-y^2}$ superconductivity) at hopping
$-t'=0.3$ and critical scale $\Omega_{\mtin{C}}$. The dependence is
obtained by an approximation with step functions characterized by $m=5$, $n=14$, and
$\alpha=2$ in (a), and $\alpha=3$ in (b). Although the peak in (b)
is very sharp, the small segments chosen at small momenta 
describe it by
more than one constant. \label{fig:BosonPropMomentum}}
\end{figure}

As described in Section \ref{sec:patching} we approximate the momentum
dependence of the boson propagators by step functions, which are constant on segments in momentum space. Even for huge segments with
$m=1$ and $n=4$ qualitatively consistent results are obtained,
although the crossover values of $-t'$ between different
instabilities as well as the critical scale are shifted compared to
Figure \ref{fig:OmegaKrit}, where we have used $m=5$ and $n=14$.
For small and big $-t'$ the critical scale shows little dependence
on different segment sizes, especially on the angular resolution.
However, in the crossover region between superconductivity and
ferromagnetism for intermediate $-t'$ the angular dependence becomes
important, although the qualitative features of Figure
\ref{fig:OmegaKrit} remain unchanged. Plotting the momentum
dependence obtained in the flow, it can be seen that only the boson
propagators $B^{(\hat{\pi})}$ away from small $-t'$ have a
significant angular dependence. Those boson propagators have four
identical maxima that move away from $\mathbf{l}=0$ as $-t'$ is
increased. In an angular sector that contains a maximum, they
slowly increase on a high plateau for small increasing radial
momentum. For higher radial momenta in the same sector they fall off
rapidly after they passed the maximum, see also Figure
\ref{fig:BosonPropWinkel}(a) at the end of this section. On the
other hand, in between angular sectors that contain a maximum, there
are sectors where the value of the boson propagator decreases from
$\mathbf{l}=0$ on with increasing radial momenta. Consider for
example Figure \ref{fig:BosonPropMomentum}(a), where the
antiferromagnetic boson propagator
$M_{\mtin{S}\mtin{S}}^{(\hat{\pi})}$ is plotted at scale
$\Omega_{\mtin{C}}$ and parameter $-t'=0.3$. At this hopping the
symmetric state is unstable towards $d_{x^2-y^2}$ superconductivity,
which is induced by the antiferromagnetic boson propagator.
Neglecting the angular momentum dependence of
$M_{\mtin{S}\mtin{S}}^{(\hat{\pi})}$ results in a slightly higher
$\Omega_{\mtin{C}}$ for superconductivity.

Generally, the magnetic and forward scattering boson propagators have sizeable contributions for a wide range of radial momenta, while the superconducting boson propagator $D_{11}$ has a very
strong peak at $\mathbf{l}=0$, especially in the region where
superconductivity is dominant.
This can be seen in Figure
\ref{fig:BosonPropMomentum}(b) where $D_{11}$ is plotted at
$\Omega_{\mtin{C}}$ and hopping $-t'=0.3$. In order to adapt to
these different momentum scales we have chosen
$\alpha_{\mtin{SC}}=3$ for all superconducting boson propagators and
$\alpha_{\mtin{MK}}=2$ for all magnetic and forward scattering boson
propagators.

Now we discuss Figure \ref{fig:OmegaKrit}. For small $-t'$ the boson
propagator $M^{(\hat{\pi})}_{\mtin{S}\mtin{S}}(\mathbf{l})$ at
$\mathbf{l}=0$ grows strongest, indicating an instability towards an
antiferromagnetic phase. Since the flow is stopped at a relatively
high scale $\Omega_{\mtin{C}}\sim 0.1$, perfect nesting--like
effects are not restricted to $-t'=0$. If $-t'$ is increased between
$0.08<-t'<0.23$ the antiferromagnetic boson propagator is still the
leading term. However its maximum is away from $\mathbf{l}=0$, that
is, it has four identical maxima, similar to Figure
\ref{fig:BosonPropMomentum}(a). This reflects a tendency to
incommensurate antiferromagnetic order, indicated by open squares in
Figure \ref{fig:OmegaKrit}.

\begin{figure}[htb]
   \begin{minipage}[r]{0.5\textwidth}
   \begin{center}
      \includegraphics[width=0.98\textwidth]{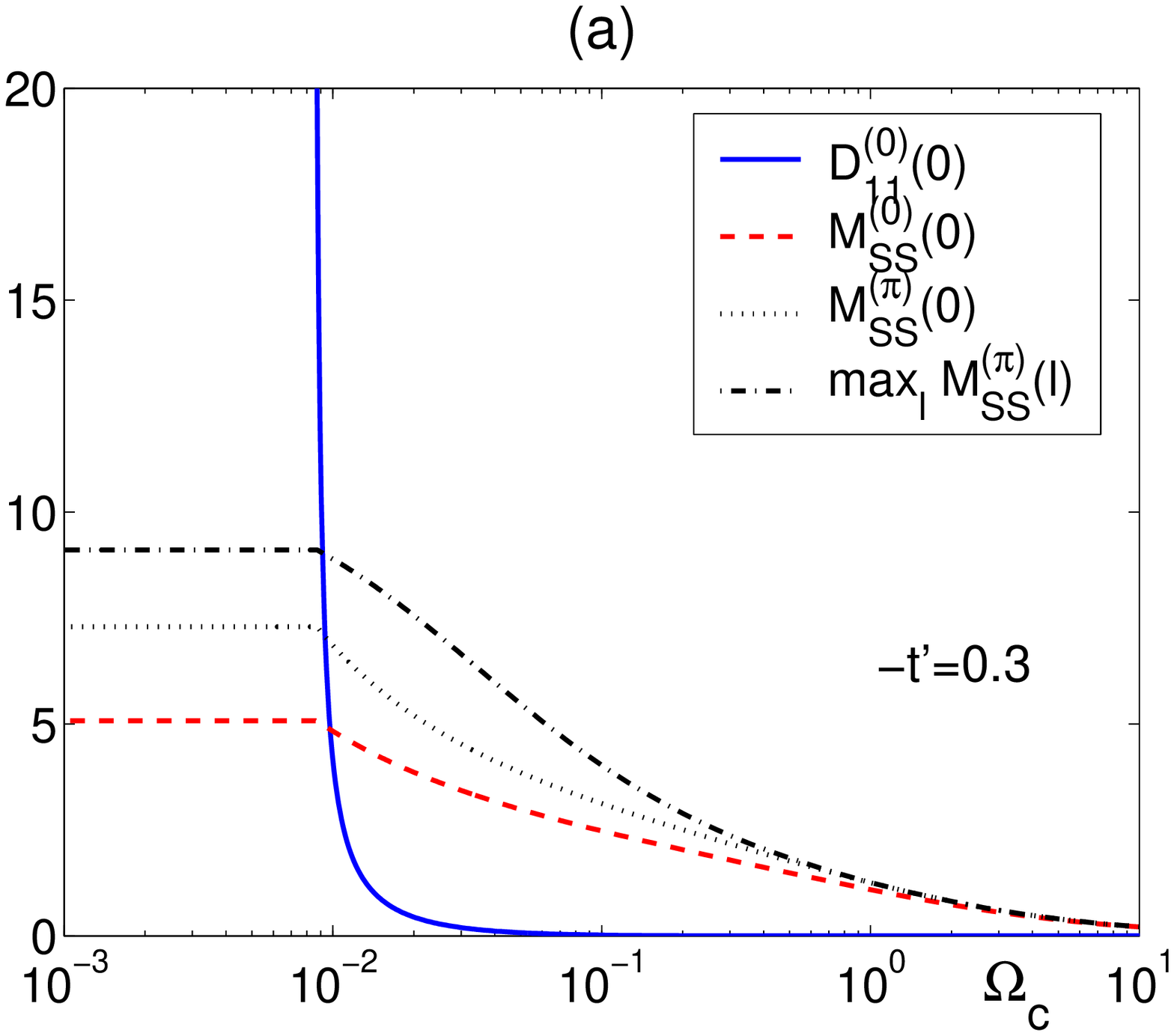}
   \end{center}
   \end{minipage}
   \begin{minipage}[c]{0.5\textwidth}
      \includegraphics[width=0.98\textwidth]{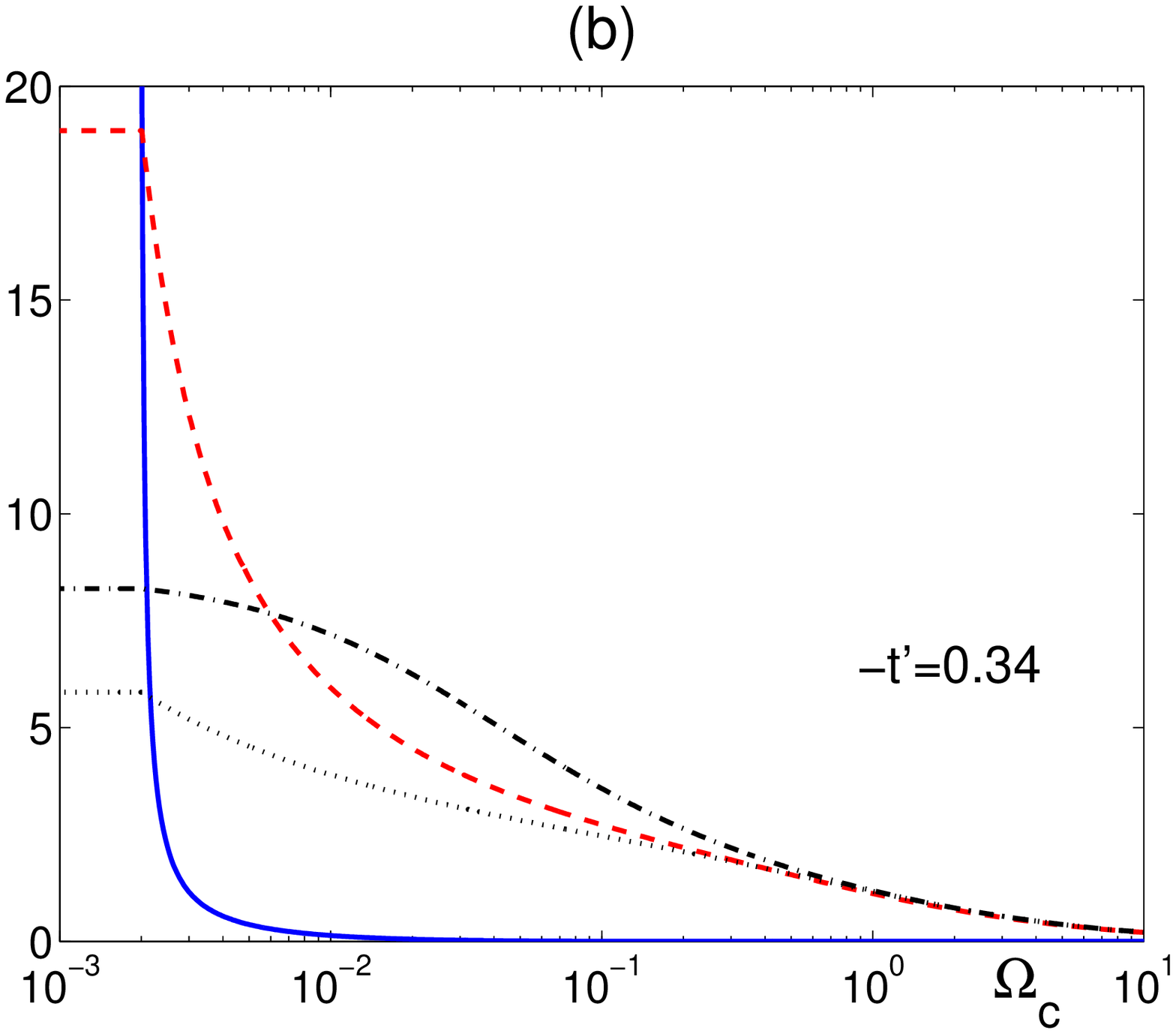}
   \end{minipage}
   \caption{The dominant flows for two different values of $-t'$ in the
   superconducting region. The peak of the boson propagator $D_{11}^{(0)}$
   at momentum zero evolves slowly but reaches the value 20 first. According to the definition of the critical scale $\Omega_{\mtin{C}}$, the  
   flow is stopped and all boson propagators are manually put constant for scales below $\Omega_{\mtin{C}}$.
   In (a) $-t'=0.3$ is well inside the superconducting region
   while in (b) $-t'=0.34$ is the crossover to the ferromagnetic
   region. Here both instabilities grow strongly. Their mutual
   suppression results in a lower critical scale. In both figures
   antiferromagnetic fluctuations are dominant for intermediate
   scales, inducing $d_{x^2-y^2}$ superconductivity.
   \label{fig:FlowExample}}
\end{figure}

\begin{figure}[htb]
   \begin{minipage}[r]{0.5\textwidth}
   \begin{center}
      \includegraphics[width=0.98\textwidth]{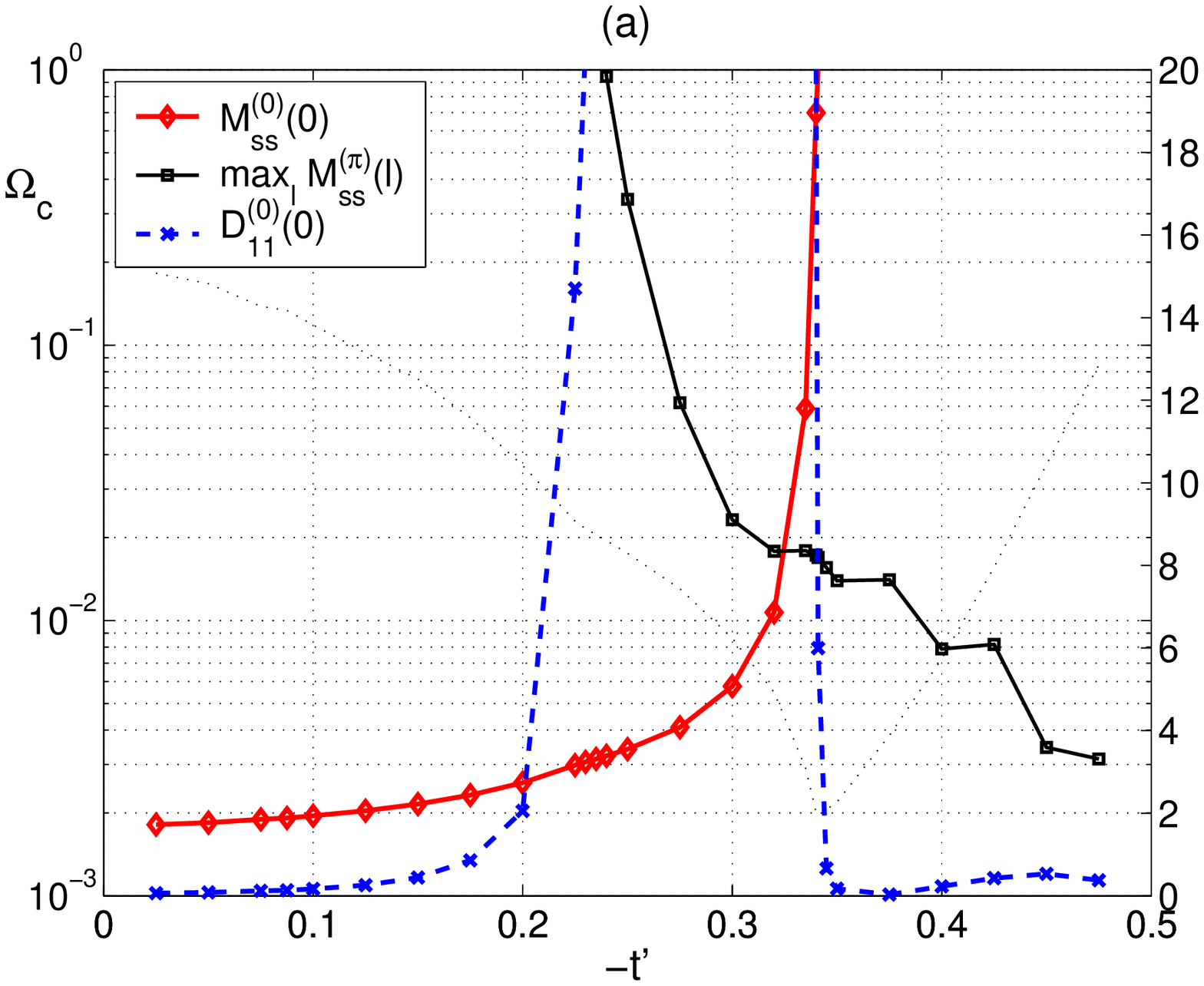}
   \end{center}
   \end{minipage}
   \begin{minipage}[c]{0.5\textwidth}
      \includegraphics[width=0.98\textwidth]{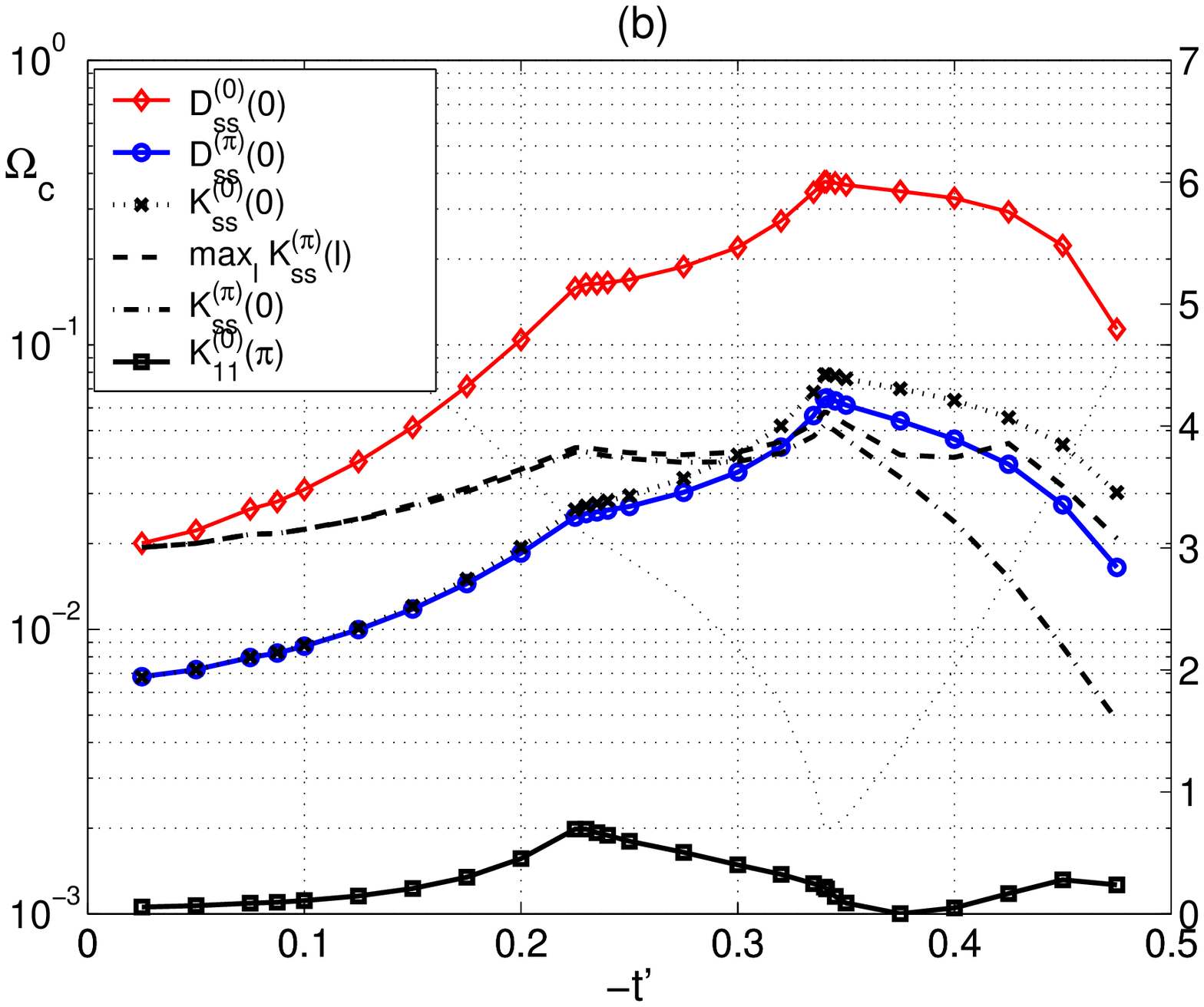}
   \end{minipage}
   \caption{Overview of the maximal value of the boson propagators at the critical scale,
   which is plotted as a thin dotted line. In (a) the three
   dominant instabilities are plotted, that is antiferromagnetism,
   $d_{x^2-y^2}$ superconductivity, and ferromagnetism. The maxima of some subleading boson propagators are plotted in (b). The kinks in the plot indicate the
   crossovers to another dominant instability. For large $-t'$ the
   discretization scheme for the momentum dependence
   becomes insufficient for the antiferromagnetic boson
   propagator, whose four maxima move towards the edge of its
   support. Due to the large segment sizes there, $\max_{\mathbf{l}}
   M_{\mtin{S}\mtin{S}}^{(\hat{\pi})}(\mathbf{l})$ jumps
   discontinuously if the maxima leave the support and another segment
   contains the new maxima, compare Figure \ref{fig:BosonPropMomentum}.}
   \label{fig:OverviewBosonprop}
\end{figure}

For $-t'\in[0.23,0.34]$ the leading term is the
superconducting boson propagator $D_{11}^{(0)}(\mathbf{l})$ at
$\mathbf{l}=0$, which represents $d_{x^2-y^2}$--wave
superconductivity. As discussed in Section
\ref{sec:superconductivity}, it is induced by antiferromagnetic
correlations. In Figure \ref{fig:FlowExample}(a) the flow of
$D_{11}^{(0)}(0)$ is plotted for $-t'=0.3$ in dependence on the
scale. First incommensurate antiferromagnetic correlations dominate
and $D_{11}^{(0)}(0)$ is enhanced by them but stays small down to
low scales. When a significant superconducting coupling is reached,
the direct graph $\sim D_{11}^{(0)}(0)^2
\dot{\Phi}_{\mtin{pp}}^{11}(0)$ contributes strongly such that
$D_{11}^{(0)}(0)$ increases very fast and becomes the dominant
coupling. The crossover between the antiferromagnetic and the
$d_{x^2-y^2}$ superconducting region is not sharp in the scheme
applied here. Around $-t'=0.23$ both tendencies grow very strongly,
making it difficult to judge numerically which is the dominant
instability. The overlap of the two regions in Figure \ref{fig:OverviewBosonprop}(a) corresponds to the saddle point region found earlier in \cite{Umklapp}.

As $-t'$ is increased the
antiferromagnetic correlation, indicated by
$|\Phi_{\mtin{ph}}(\hat{\pi})|$, decreases, while ferromagnetism is
enhanced because $|\Phi_{\mtin{ph}}(0)|$ becomes bigger. Both affect
the critical scale in the superconducting region, which is
continuously dropping, especially when $-t'$ gets close to the
ferromagnetic region. This indicates that $d_{x^2-y^2}$
superconductivity is not only induced by antiferromagnetic
correlations but also suppressed by ferromagnetic tendencies.

\begin{figure}[htb]
   \begin{minipage}[r]{0.465\textwidth}
   \begin{center}
      \includegraphics[width=0.98\textwidth]{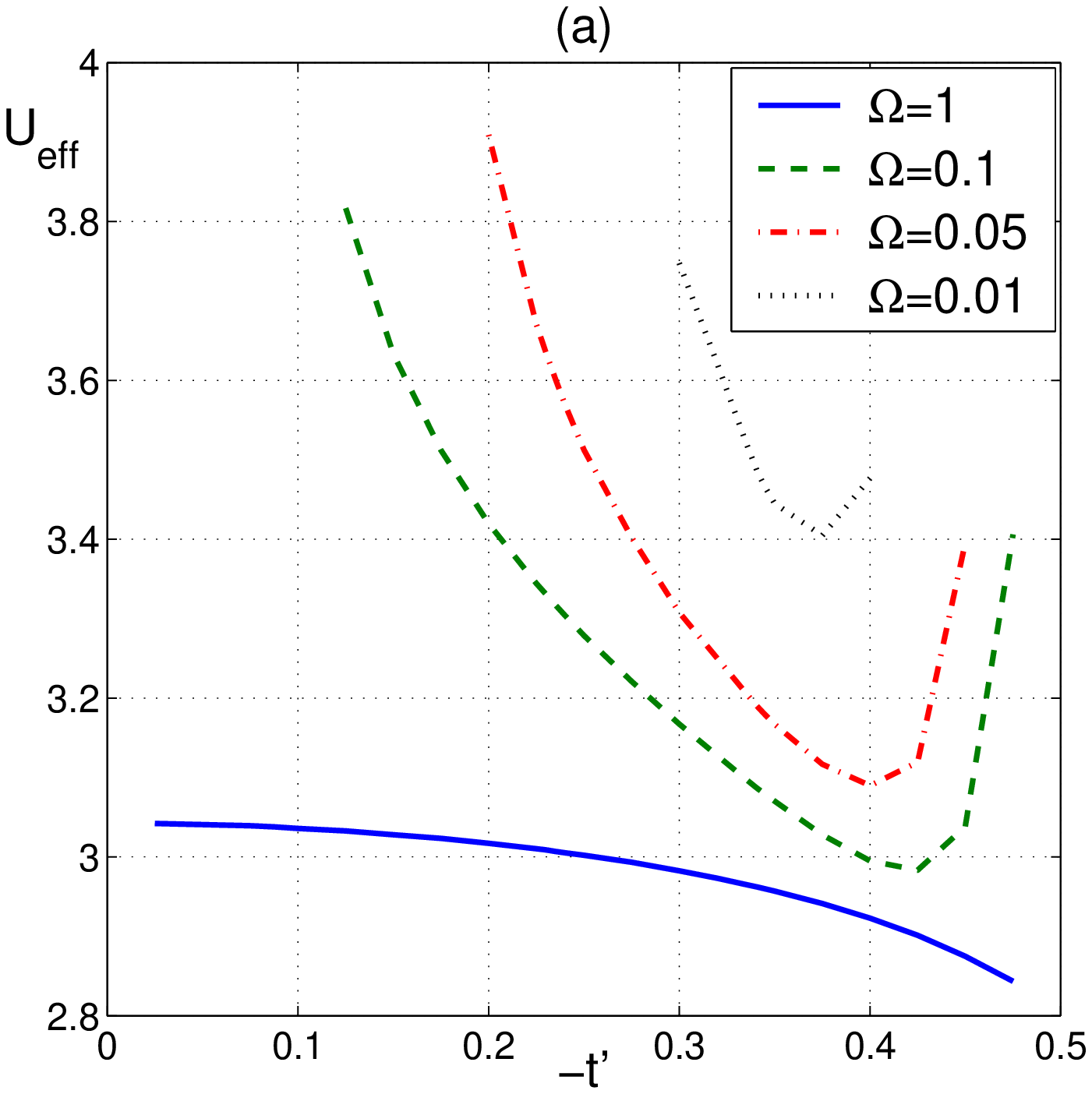}
   \end{center}
   \end{minipage}
   \begin{minipage}[c]{0.535\textwidth}
      \includegraphics[width=0.98\textwidth]{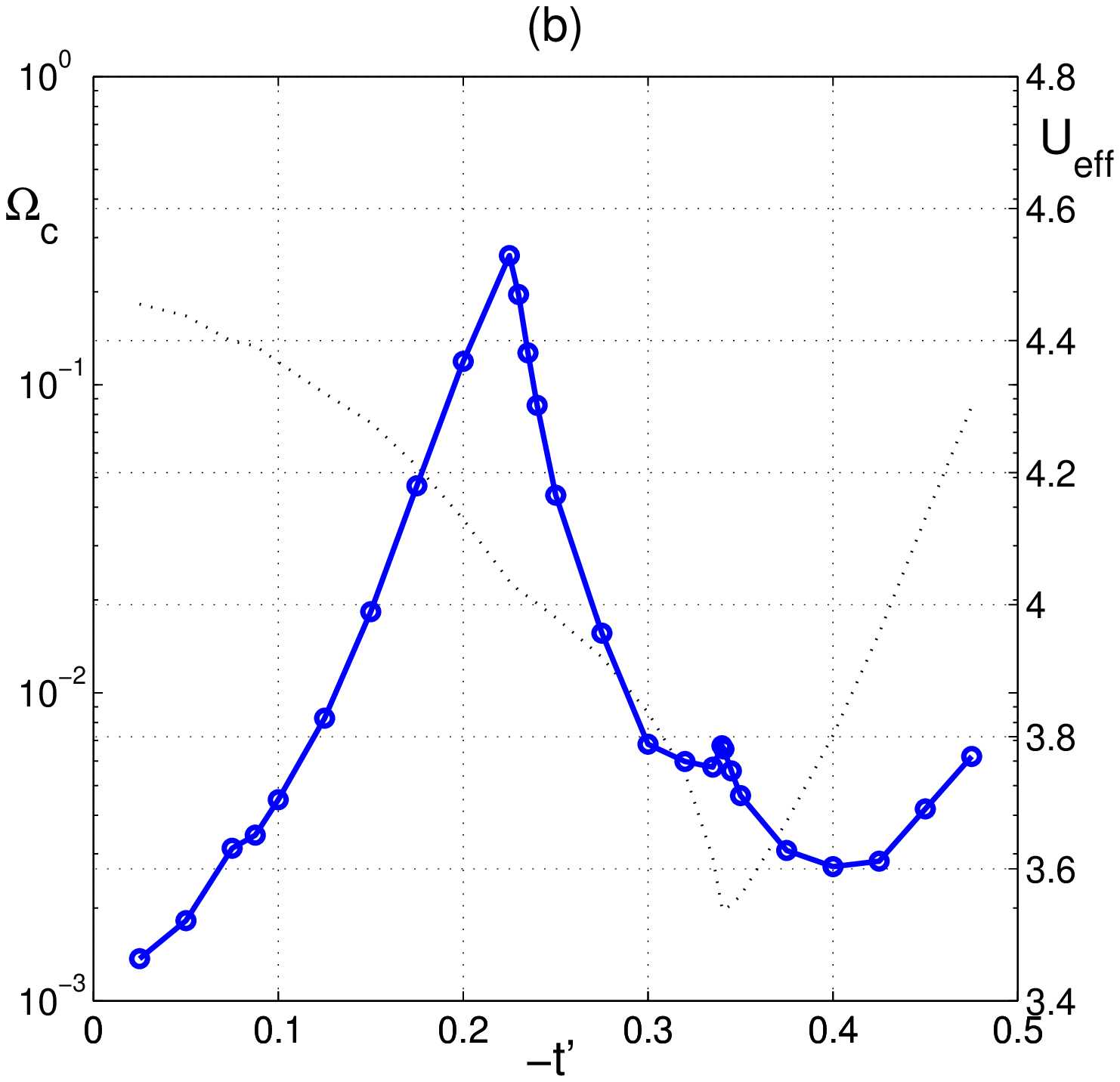}
   \end{minipage}
   \caption{The effective on--site coupling $U_{\mtin{eff}}$ is plotted in dependence of $-t'$ for different scales in (a). The curves to different values of $\Omega$ are drawn only over the interval where that $\Omega$ is still bigger than the critical scale $\Omega_{\mtin{C}}$. In (b)   $U_{\mtin{eff}}$ is evaluated at the critical scale, which is plotted as a dotted line for comparison. The kinks in (b) indicate the crossovers of distinct instabilities. }
   \label{fig:onsiteU}
\end{figure}

\begin{figure}[htb]
   \begin{minipage}[r]{0.5\textwidth}
   \begin{center}
      \includegraphics[width=0.98\textwidth]{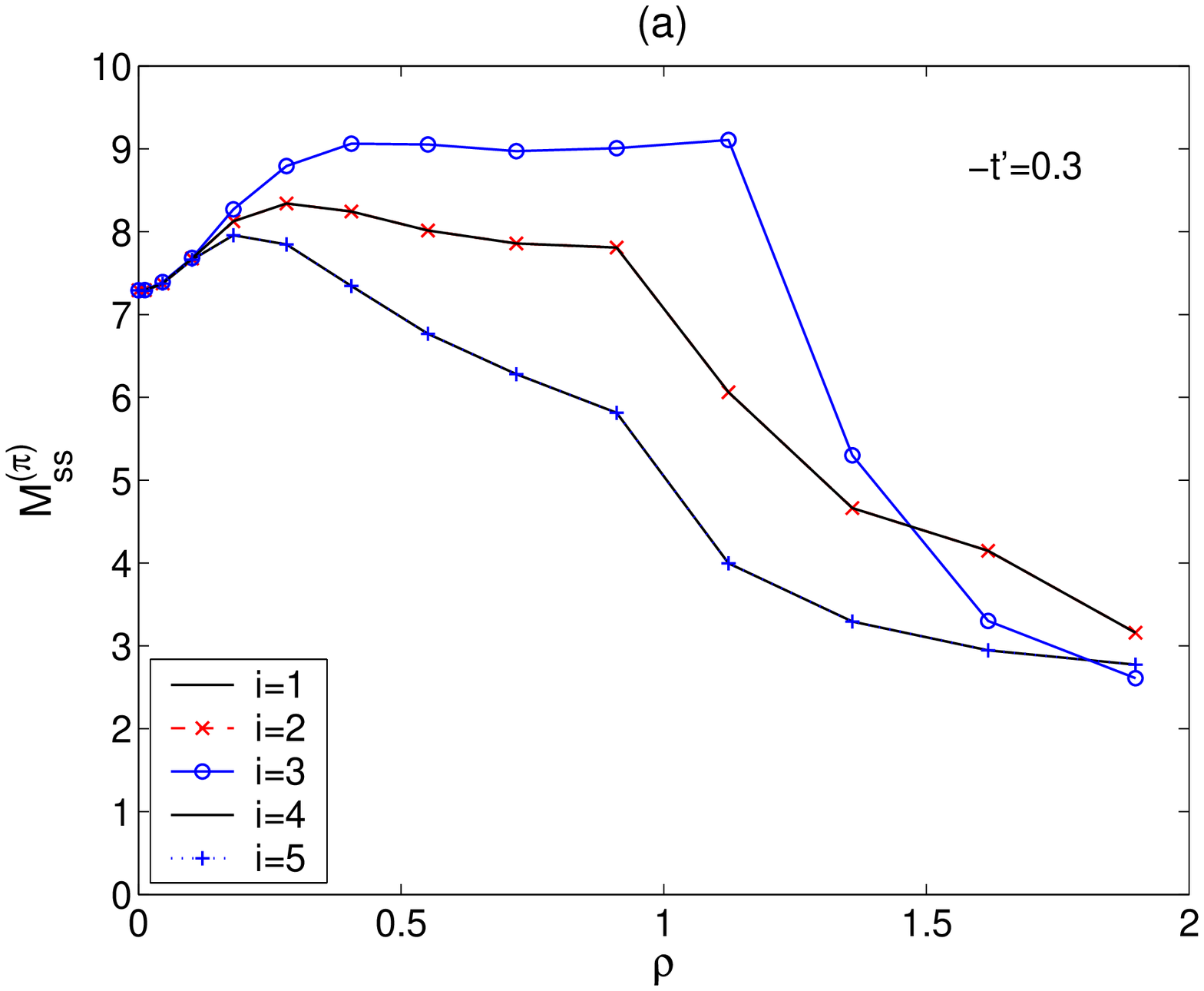}
   \end{center}
   \end{minipage}
   \begin{minipage}[c]{0.5\textwidth}
      \includegraphics[width=0.98\textwidth]{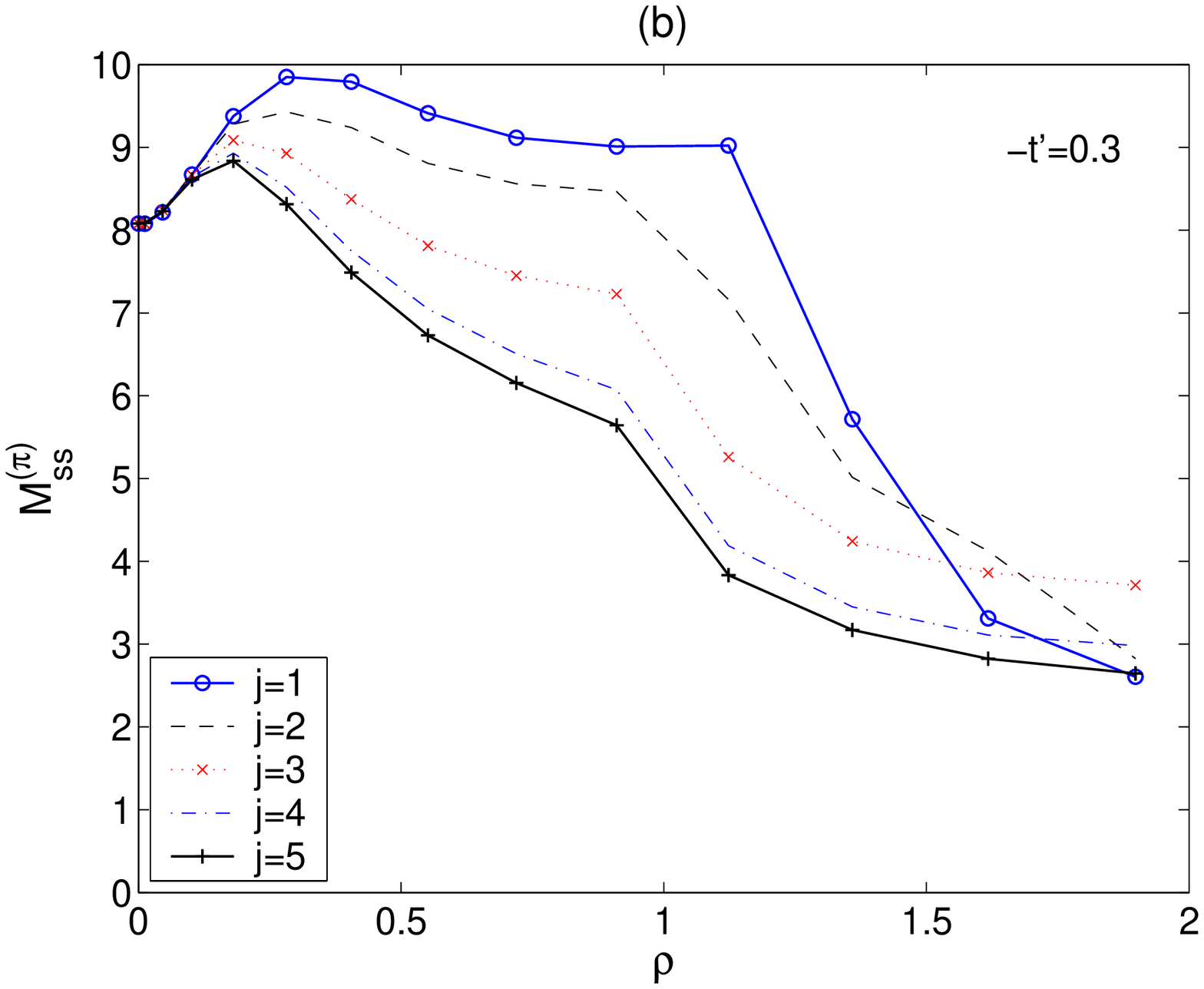}
   \end{minipage}
   \caption{The antiferromagnetic boson propagator
   $M_{\mtin{SS}}^{(\hat{\pi})}$ plotted over radial momenta for
   different angular sectors centered at $\phi=\arctan
   \frac{l_y}{l_x}$. For the first numerical calculation
   shown in (a) the centers of the angular sectors take the values
   $\phi_i=-\frac{\pi}{4} + (i-\frac 12 )\frac{\pi}{10}$ and for the computation with an offdiagonal boson propagator in
   (b) the angular sectors are centered at
   $\phi_j=(j-\frac 12 )\frac{\pi}{20}\,.$}
   \label{fig:BosonPropWinkel}
\end{figure}

For high $-t'>0.34$ the leading instability is ferromagnetism,
represented by the dominant flow of
$M^{(0)}_{\mtin{S}\mtin{S}}(\mathbf{l}=0)$. The critical scale rises
again as $-t'$ increases. However, the critical scale is still about
one order of magnitude lower than the Stoner criterion suggests. The
Stoner criterion for ferromagnetism is obtained by neglecting the
particle--particle channel and given by $U|\Phi_{\mtin{ph}}(0)|=1$.
Therefore superconducting fluctuations suppress ferromagnetism as
well. However, outside the $d$--wave superconductivity region the
superconducting $d$--wave boson propagator
$D_{11}^{(0)}(\mathbf{l})$ remains small, see Figure
\ref{fig:OverviewBosonprop}(a) for its maximum at $\mathbf{l}=0$. A
significant contribution to this suppression comes from $s$--wave
superconductivity $D_{\mtin{S}\mtin{S}}^{(0)}$ and
$D_{\mtin{S}\mtin{S}}^{(\hat{\pi})}$, so--called screening, compare Figure \ref{fig:onsiteU}.

In summary, ferromagnetism and $d$--wave superconductivity are
competing instabilities of the $(t,t')$--Hubbard model for
relatively large $-t'$ at Van Hove filling. In the calculation
presented here the crossover between superconductivity and
ferromagnetism takes place at $-t'=0.34$ which is in good agreement
with $-t'=0.33$ found with the temperature RG flow
\cite{TemperatureFlow} and the two--particle self--consistent Monte
Carlo approach \cite{Tremblay}. This is also roughly the value of
$-t'$ where $\Phi_{\mtin{ph}}(0)=\Phi_{\mtin{ph}}(\hat{\pi})$.
However, compared to the temperature RG flow, the suppression
between these competing tendencies is weaker in the approximation
used here.  
Although for low and high $-t'$ the critical scale
$\Omega_{\mtin{C}}$ and the critical temperature from
\cite{TemperatureFlow} is roughly the same, the temperature RG flow
finds  critical scales that are two orders of magnitude lower in the
crossover region between superconductivity and ferromagnetism.
Especially, the results obtained here do not suggest a quantum
critical point separating these two instabilities in the phase
diagram. Instead, in the crossover region both processes grow
strongly like in the crossover between antiferromagnetism and
superconductivity. However, the crossover is much sharper as can be
seen in Figure \ref{fig:OverviewBosonprop}(a). The superconducting
boson propagator $D_{11}^{(0)}$ remains small as soon as $-t'$
enters the ferromagnetic region, while in the antiferromagnetic
region superconducting tendencies persist, such that this crossover
region is broader. 

At present, it is not clear to us where this difference to \cite{TemperatureFlow} comes from, as both schemes contain rather different approximations. The RG equations for the boson propagators imply
that their diagonal parts are monotonically increasing in the flow.
Already a small inexactness, such as the omission of the remainder
terms or the discretization of the momentum dependence, can prevent
an exact cancellation on the right hand side of the flow equation for
$D_{11}^{(0)}$ and $M_{\mtin{S}\mtin{S}}^{(0)}$. On the other
hand, the projection to the Fermi surface, used in
\cite{TemperatureFlow} but not here, is expected to enhance
superconducting tendencies. This could possibly lead to a stronger
suppression. 
Furthermore, the $\Omega$--scheme used here allows a clearer definition of the initial conditions than the temperature scheme.
It would be interesting to perform an $N$--patch analysis with the proposed
$\Omega$--regularization (\ref{eq:OmegaCutoff}).

So far we have not discussed all boson propagators computed in the
flow. The $s$--wave forward scattering boson propagators
$K_{\mtin{SS}}^{(0)}$ and $K_{\mtin{SS}}^{(\hat{\pi})}$ and the
$s$--wave superconducting boson propagators $D_{\mtin{SS}}^{(0)}$
and $D_{\mtin{SS}}^{(\hat{\pi})}$ cannot become singular since they are dominated by the initial local repulsion $U$. In Figure
\ref{fig:OverviewBosonprop}(b) their final values are plotted over
hopping $-t'$. They are biggest in the crossover region between
superconductivity and ferromagnetism, indicating the amount of
screening and the low critical scale.  The boson propagator
$K_{11}^{(0)}$, representing a possible Pomeranchuk instability, is
induced by magnetic correlations in the flow. However, it is not
dominant in the
parameter region considered here.
The other boson propagators $D_{11}^{(\hat{\pi})}$,
$M_{11}^{(0)}$, $M_{11}^{(\hat{\pi})}$, and $K_{11}^{(\hat{\pi})}$
remain smaller than 0.05 and hence are irrelevant here. Therefore, in order to further reduce computing cost, they could be omitted. In fact, only the boson propagators
$D_{11}^{(\hat{\pi})}$, $M_{11}^{(\hat{\pi})}$, and
$K_{11}^{(\hat{\pi})}$ do not satisfy the symmetry
$B^{(a)}(l_x,l_y)=B^{(a)}(l_x,-l_y)$. If they are not computed in
the flow, the angular segments only need to cover the interval
$[0,\frac{\pi}{4}]$. This would halve the number of step functions necessary for the discretization of the momentum dependence described
in Section \ref{sec:patching}.

We have already mentoined that the effective on--site interaction changes although the initial on--site repulsion $U$ of the Hubbard model is kept constant in the parametrization of the vertex function (\ref{eq:decompositionvertexfunction}). Projecting to the on--site part gives the scale--dependent effective on--site coupling
\begin{align*}
U_{\mtin{eff}}= U+ \int_{|l_x|+|l_y|\le \pi} \frac{\mathrm{d}^2 \mathbf{l}}{(2\pi)^2}\sum_{a=0,\hat{\pi}} \left[ -D_{\mtin{S}\mtin{S}}^{(a)}(\mathbf{l}) +\frac 32 M_{\mtin{S}\mtin{S}}^{(a)}(\mathbf{l}) - \frac 12 K_{\mtin{S}\mtin{S}}^{(a)}(\mathbf{l}) \right]\, ,
\end{align*}
which is plotted in Figure \ref{fig:onsiteU} over hopping $-t'$. Although the critical scale is lowest in between the superconductor and the ferromagnet, the effective on--site coupling has not increased much in comparison to higher scales. This indicates that the amount of screening, especially in the crossover region between superconductivity and ferromagnetism, is substantial.

In a second numerical computation we study the influence of
offdiagonal boson propagators. In the superconducting channel we
consider the boson propagators $D_{\mtin{SS}}^{(0)}$,
$D_{\mtin{SS}}^{(\hat{\pi})}$, $D_{11}^{(0)}$, and
$D_{1\mtin{S}}^{(0)}$. The first two are important to capture
screening effects, the third describes $d_{x^2-y^2}$
superconductivity and the last is an offdiagonal boson propagator
that couples $s$-- and $d$--wave superconductivity. In the magnetic
and forward scattering channel we only take into account the local
$s$--wave boson propagators $M_{\mtin{SS}}^{(0)}$,
$M_{\mtin{SS}}^{(\hat{\pi})}$, $K_{\mtin{SS}}^{(0)}$, and
$K_{\mtin{SS}}^{(\hat{\pi})}$. Compared to the previous numerical
calculation, apart from the irrelevant boson propagators $D_{11}^{(\hat{\pi})}$,
$M_{11}^{(0)}$, $M_{11}^{(\hat{\pi})}$, and $K_{11}^{(\hat{\pi})}$, only the boson propagator
$K_{11}^{(0)}$, which indicates a possible Pomeranchuk instability,
is omitted. However, $K_{11}^{(0)}$ remained relatively small in the
first computation, see Figure \ref{fig:OverviewBosonprop}. Note that
for this choice of boson propagators there are no more offdiagonal
terms than the one included.

The flow equations imply that the diagonal boson propagators obey
the additional symmetry $B^{(a)}(l_x,l_y) = B^{(a)}(l_x,-l_y)$. The
offdiagonal boson propagator $D_{1\mtin{S}}^{(0)}$ obeys this
symmetry as well and is also symmetric under $\mathbf{l}\to
-\mathbf{l}$. However, it is antisymmetric under exchange of
$l_x\leftrightarrow l_y$. Therefore, compared to the choice of step functions
described in Section \ref{sec:patching}, we only need to consider segments with an
angle $\phi=\arctan \frac{l_y}{l_x}$ in the interval
$[0,\frac{\pi}{4}]$. Since we still use the parameter values $m=5$
and $n=14$, the angular resolution is higher in this second
numerical computation. Again we computed the RG flow of the boson
propagators for coupling $U=3$ at temperature zero and Van Hove
filling for various $-t'\in(0,\frac12)$. 
Since no major process
was omitted compared to the first numerical computation, no
significant change could be observed in the figures presented.
Furthermore, the maximum of the offdiagonal boson propagator $D_{1\mtin{S}}^{(0)}$ remains
smaller than $0.3$ for $-t'\ge 0.37$ and smaller than $0.03$ for $-t'<0.37$. The offdiagonal boson propagator that
couples $s$-- and $d$--wave superconductivity can therefore be neglected in
good approximation.

The only difference in the two different
numerical calculations is found in the shape of the
antiferromagnetic boson propagator $M_{\mtin{SS}}^{(\hat{\pi})}$ for
relatively high $-t'$. We already commented on the insufficient
description of large radial momenta
of the antiferromagnetic boson propagator $M_{\mtin{SS}}^{(\hat{\pi})}$, which is
plotted in Figure \ref{fig:BosonPropWinkel} at $-t'=0.3$ in dependence on $\rho$ in different angular sectors, in (a)
for the first and in (b) for the second numerical computation of
this section. Note that the angular degeneracy in (a) is due to the
same number of angular segments in the larger angular interval $[-\frac{\pi}{4},\frac{\pi}{4}]$.
The major difference between both figures is that in (b) the boson
propagator has its maximal value at smaller radial momentum $\rho$.
This can play a role in the search for an analytic parametrization
of the antiferromagnetic boson propagator, although the plateau in
the angular sector $\phi=0$ for intermediate $\rho$ seems to be its
most important feature. However, note that the angle $\phi=0$ is not
a center of an angular sector in (b). The maximal value of the boson
propagator is found in the nearest angular segment, which contains
$\phi=0$ but is centered at $\phi=\frac{\pi}{40}$. This could also
be a partial explanation of the observed difference.

\section{Conclusion}

We have presented a novel parametrization of the one--loop
one--particle irreducible RG equation for the four--point function and applied it to the two--dimensional Hubbard model using a novel RG regularization scheme.
The parametrization is based on the idea to separate dominant interactions from
irrelevant remainders, where the latter do not influence the leading
instabilities of the flow. This idea is implemented by writing the
effective two--fermion interaction as a sum of dominant terms where fermion bilinears
interact via exchange bosons. Each such term consists of a boson propagator and boson--two--fermion vertex functions. From the singular momentum
structure of the RG equation we identify three different channels in
the interaction. In these channels the boson--fermion vertex
function is expanded in scale--independent form factors. We argued
that only a few terms are needed in the expansions to describe the
qualitative structure of the flow. The tails of the expansions are
the above mentioned remainder terms, which are neglected. Although the scale
dependent coefficient functions of the expansions are called boson
propagators, they are kept as fermionic interaction terms. Their flow is computed by suitable projections of
the fermionic RG equation.

The application of this method to the $(t,t')$--Hubbard model at Van
Hove filling and temperature zero shows indeed that the essential
qualitative structure of the one--loop RG is preserved. We are able
to reproduce the leading weak coupling instabilities of the model,
as they were found in previous RG studies using $N$--patch schemes.
In particular, our results are in good qualitative agreement with
the temperature RG flow \cite{TemperatureFlow}. We use a new
frequency regularization, which like the temperature regularization
does not artificially suppress ferromagnetism, but which allows to
state the initial conditions of the RG flow more clearly. For an
initial repulsive on--site interaction $U=3$ we find three distinct
regions, characterized by the leading weak coupling instability and
depending on next to nearest neighbor hopping, see Figure
\ref{fig:OmegaKrit}. For small next to nearest neighbor hopping
$-t'$ antiferromagnetism dominates. As was found in the $N$--patch
schemes, antiferromagnetic correlations induce a $d_{x^2-y^2}$--wave
superconductivity instability, which becomes dominant for
intermediate $-t'$. For high $-t'$ the leading instability is
ferromagnetism. In between the $d_{x^2-y^2}$--wave superconducting
and the ferromagnetic region the critical scale drops by two orders
of magnitude. We argue that this is due to the mutual suppression of
these opposing correlations. Compared to \cite{TemperatureFlow},
however, we find that this suppression is weaker here. In
particular, our result does not suggest a quantum critical point
between the two regions. Because we have not been able to trace this discrepancy to certain scattering processes, we regard this matter as open and leave it for further study. 

The proposed method is in general not restricted to temperature zero
and Van Hove filling. Finite temperature only affects the
calculation of the one--loop fermion bubbles, which then becomes
more involved. Away from Van Hove filling triplet superconductivity
plays a role for high $-t'$ \cite{TemperatureFlow}. The flow
equations stated in Section \ref{sec:RGequations} can account for
that. However, it is not clear that a single form factor suffices to
describe dominant triplet superconductivity. The results of \cite{TemperatureFlow} suggest that at least three different triplet form
factors would have to be taken into account.

The benefit of the proposed parametrization of the vertex function
is a reduction of the complexity of the full one--loop flow to some
dominant terms. We believe that separating leading from subleading
processes will help to gain further insight into the structure of
the one--loop RG. However, we have not performed a detailed 
analysis of the remainder.
The comparison with
previous $N$--patch studies suggests, nevertheless, that we did
capture the most important processes. A rigorous proof is
particularly complicated at Van Hove filling due to the strong
mixing of the particle--particle and the particle--hole channels. We
gave clear arguments for determining the dominant terms in case of a
regular and curved Fermi surface.

Practically, this reduction of complexity results in lower computing
cost compared to previous $N$--patch schemes. In the numerical
implementation the momentum dependence of the boson propagators is
discretized using step functions. The choice of segments, where the
step functions are constant, can be guided by the form of the
one--loop bubbles. This allows a more precise discretization than
the general patching of the vertex function in $N$--patch schemes.
In particular, no momentum dependence is projected onto the Fermi
surface. Furthermore, the proposed decomposition gives possibly way
to further improvement. If the momentum dependence of the one--loop
bubbles can be parameterized in an analytical form, then it should
be possible to extract a functional parametrization of the boson
propagators from the flow equations, at least for small momenta.
Deviations for large momenta, away from the maxima of the boson
propagators, could be subject to another negligible remainder.
However, the momentum dependence of the bubbles is difficult to
describe since a naive power expansion is not sufficient.
Nevertheless, while in the numerical calculation we have neglected
the frequency dependence of the boson propagators, a similar
procedure for small frequencies could take into account at least
part of the frequency dependence of the vertex function.

The proposed decomposition of the effective two--fermion interaction
is of a form that suggests the decoupling of the fermion bilinears
via multiple Hubbard Stratonovich transformations. The ambiguity of
introducing boson fields is not completely removed but at least
reduced. This allows to continue the RG flow into the symmetry
broken phase in a (partially) bosonized form.

We acknowledge financial support from DFG grant SA 1362/1--2 and
DFG research group FOR 723.
\bibliography{decomposition}
\end{document}